\documentclass[twocolumn,tighten]{aastex63}
\pdfoutput=1 
\usepackage{amsmath,amstext}
\usepackage{apjfonts} 
\usepackage{afterpage}
\usepackage{xcolor}
\usepackage{threeparttable}
\usepackage{lineno}
\usepackage{graphicx}
\usepackage{epsfig}

\usepackage{booktabs}
\usepackage{multirow}
\usepackage{float}
\usepackage{mathrsfs}
\DeclareUnicodeCharacter{1E11}{\k{a}}
\setwatermarkfontsize{100px} 

\shortauthors{Satapathy, Psaltis, \& \"Ozel}
\begin{document}


\title{The Origin of the Slow-to-Alfv\'en Wave Cascade Power Ratio and its Implications for Particle Heating in Accretion Flows}
\shorttitle{}

\author{Kaushik Satapathy}
\affiliation{School of Physics, Georgia Institute of Technology, 837 State St NW, Atlanta, GA 30332, USA}
\affiliation{Department of Physics, University of Arizona, 1118 E. Fourth Street, Tucson, AZ 85721}

\author{Dimitrios Psaltis}
\affiliation{School of Physics, Georgia Institute of Technology, 837 State St NW, Atlanta, GA 30332, USA}

\author{Feryal \"Ozel}
\affiliation{School of Physics, Georgia Institute of Technology, 837 State St NW, Atlanta, GA 30332, USA}

\begin{abstract}
The partition of turbulent heating between ions and electrons in radiatively inefficient accretion flows plays a crucial role in determining the observational appearance of accreting black holes. Modeling this partition is, however, a challenging problem because of the large scale separation between the macroscopic scales at which energy is injected by turbulence and the microscopic ones at which it is dissipated into heat. Recent studies of particle heating from collisionless damping of turbulent energy have shown that the partition of energy between ions and electrons is dictated by the ratio of the energy injected into the slow and Alfv\'en wave cascades as well as the plasma $\beta$ parameter. In this paper, we study the mechanism of the injection of turbulent energy into slow- and Alfv\'en- wave cascades in magnetized shear flows. We show that this ratio depends on the particular ($r\phi$) components  of the Maxwell and Reynolds stress tensors that cause the transport of angular momentum, the shearing rate, and the orientation of the mean magnetic field with respect to the shear. We then use numerical magnetohydrodynamic shearing-box simulations with background conditions relevant to black hole accretion disks to compute the magnitudes of the stress tensors for turbulence driven by the magneto-rotational instability and derive the injection power ratio between slow and Alfv\'en wave cascades. We use these results to formulate a local subgrid model for the ion-to-electron heating ratio that depends on the macroscopic characteristics of the accretion flow. 
 
\end{abstract}


\section{Introduction}
\label{sec:Intro}

Recent Very Long Baseline Interferometric (VLBI) observations by the Event Horizon Telescope of the horizon-scale environments of the black holes at the centers of the Milky Way (Sgr A*) and M87 galaxies have offered a unique opportunity to study plasma astrophysics in low-luminosity accretion flows \citep{EHT2019e,EHT2022e}. These systems are categorized as Radiatively Inefficient Accretion Flows \citep[see][]{Narayan1995b} owing to their low luminosities and the characteristics of their spectra. They are made up of low density plasmas, with mass accretion rates typically less than $10^{-3}$ times the Eddington rate $\dot{M}_{\rm{Edd}}$. The collisional timescales between the ions and the electrons in these flows are much larger than accretion timescale, allowing for the ions and the electrons to co-exist in a two-temperature state. In addition, low densities also cause the cooling processes in these systems to be inefficient, resulting in most of the energy carried by the plasma to be advected into the black hole.

Due to the two-temperature nature of the plasma, a first-principles approach to studying accretion flows involves independently evolving the thermodynamics of the ions and the electrons, incorporating physical models for the partition of heat between the species \citep{Ressler2015,Sadowski2017}. In these accretion flows, heating mechanisms include collisionless damping due to wave-particle resonances \citep{Quataert1998,Kawazura2020}, magnetic reconnection \citep{Ball2018,Rowan2019}, and heating resulting from instabilities due to velocity-space anisotropies \citep{Sharma2007}. These processes are typically dominant at the length-scales comparable to the gyroradii of the ions and the electrons, which are much smaller than the scale of the system. As a result of the large scale separation, the physics of heating mechanisms studied at the microscopic kinetic scales need to be incorporated as sub-grid prescriptions in global simulations. 

Among these dissipation processes, magnetic reconnection is primarily dominant in localized regions of the flow that have current sheets \citep{Ball2016} and causes episodic dissipation of energy. On the other hand, velocity space anisotropies are believed to be driven by mechanisms connected to the time-evolving large scale magnetic fields. In the absence of current sheets or local mechanisms driving velocity space anisotropies, the turbulent energy cascades down to scales comparable to the ion gyroradius and undergoes collisionless damping. This is a ubiquitous mechanism of dissipation, both spatially and temporally and is expected to be the primary driver of electron and ion heating.

Recent numerical studies show that the partition of heat between ions and electrons resulting from collisionless damping of the turbulent energy \citep{Kawazura2020} depends on the plasma $\beta$ (the thermal to magnetic pressure ratio) and the ratio of the driving power of the compressive slow magnetosonic wave cascade to that of the Alfvén wave cascade ($P_{S}/P_A$). These studies have been performed in the gyrokinetic limit, where the gyromotion of particles around the mean magnetic field are averaged out. In such a limit, the compressive slow waves and the Alfvén waves in the turbulence decouple at the leading order at scales smaller than the driving scales of the turbulence \citep{Schekochihin2009}. Owing to this inertial range, the ratio $P_S/P_A$ does not only impact the microphysical partition of heat between ions and electrons but is also evident in the ion and electron temperature profiles in an accretion flow \citep{Satapathy2023}, underscoring the importance of studying the nature of the slow and Alfvén driving caused by the MRI.

The ratio of the driving power of the slow- and the Alfvén- waves, in turn, is determined by the mechanism that drives turbulence in the flow. In black hole accretion flows, the magneto-rotational instability \citep[MRI, see][]{Balbus1991} is believed to drive the turbulence and cause the outward transport of angular momentum.  The physics of the injection of energy into the slow and Alfvén wave turbulent cascades caused by the MRI has not been extensively explored. Only one study, performed by \citet{Kawazura2022}, addresses the problem numerically using reduced-MHD shearing box simulations. These authors studied the MRI driven by a near-azimuthal mean magnetic field in a Keplerian shear flow and showed that the slow-wave cascade is driven approximately at twice the power as the Alfvén-wave cascade.

In this paper, we study the slow-to-Alfvén wave driving power ratio $P_S/P_A$ and establish its relation to the large-scale properties of the turbulence. We show that the ratio $P_S/P_A$ depends on particular components of the Maxwell and Reynolds stress tensors that cause the outward transport of angular momentum in the disk, the orientation of the mean magnetic field, and the shearing rate. We then employ local shearing box simulations in presence of different mean magnetic field configurations and shearing rates relevant to accretion flows to compute the two stress tensors resultant due to the MRI and its parasitic instability in a non-linear turbulent state. Subsequently, we infer the slow-to-Alfvén driving power ratio $P_S/P_A$ for these configurations. In combination with hybrid kinetic scale simulations performed by \cite{Kawazura2020}, we formulate a local sub-grid model for the partition of turbulent heat between the ions and the electrons resultant from collisionless damping of slow- and Alfvén-waves. We also resolve a confusion in the literature regarding the compressive nature of slow waves in the reduced-MHD limit and establish that the Helmholtz decomposition is not useful in calculating the ratio $P_S/P_A$ (see Appendix \ref{Appendix A}).

In Section \ref{sec:Review}, we review the effects of the large scale turbulent driving on the dissipation scale partition of heat into the ions and electrons resultant from the collisionless damping of the turbulence. In Section \ref{sec:Equations}, we set up equations of the shearing box approximation of MHD and derive the energy equations for the slow magnetosonic and the Alfvén wave fluctuations valid to the first order in the reduced-MHD limit. In Section \ref{sec:Physics}, we discuss the physics of the injection of energy into the slow and the Alfvén wave cascades as caused by the MRI and establish the relation of the injection ratio to the Maxwell and Reynolds stress tensors. In Section \ref{sec:Simulations}, we compute the value of $P_S/P_A$ for a wide range of conditions relevant to the turbulence driven in black hole accretion flows using a suite of shearing box simulations. We summarize our findings in Section \ref{sec:Discussion}.

\section{Effects of Turbulent Driving on Collisionless Damping}
\label{sec:Review}

\begin{figure}
    \includegraphics[width=0.99\linewidth]{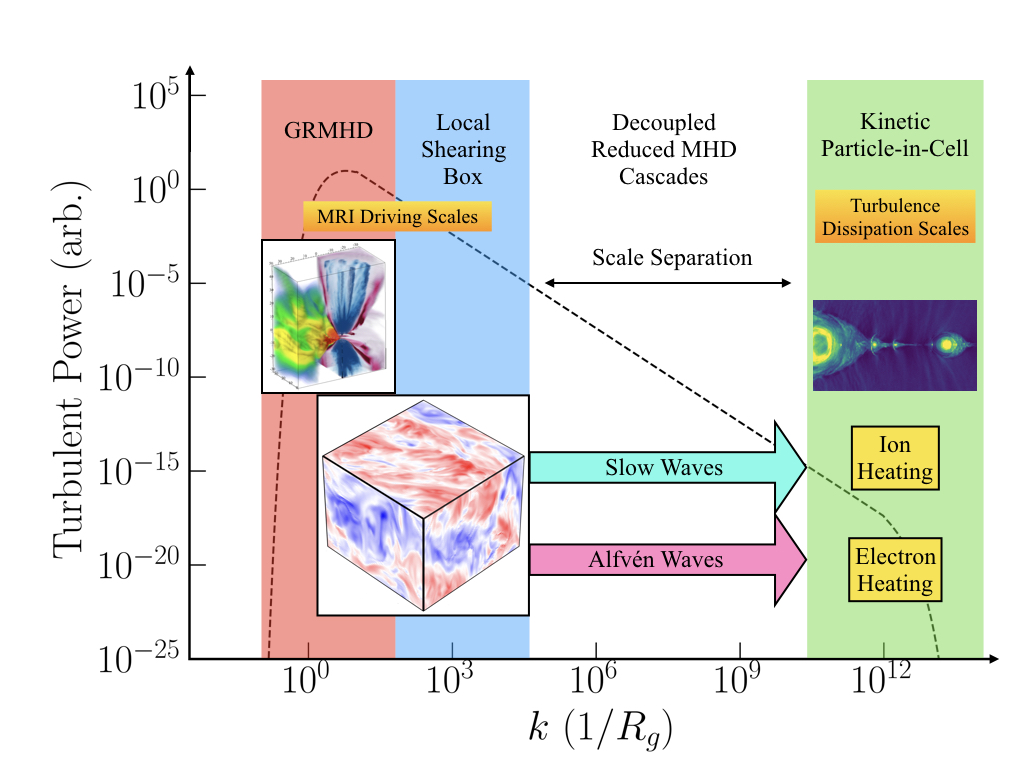}
    \caption{A representative figure of the turbulent power spectrum in black hole accretion flows (where $k$ represents the wavenumber in units of the gravitational radius $R_g$). We depict the scales numerically captured by GRMHD, local shearing box, and kinetic simulations. The large scale dynamics and global angular momentum transport is captured by GRMHD simulations. The properties of turbulence such as the power spectra and the dynamics of stresses can be studied using local shearing box approaches. Past the injection scales, the slow and Alfvén wave cascades of turbulence are energetically decoupled from each other until reaching kinetic scales compared to the ion Larmor radius. At these dissipation scales, kinetic studies help in studying the partition of heating between the ions and the electrons. }
    \label{fig:scales}
\end{figure}

The turbulence in plasmas found in radiatively inefficient accretion is believed to be strongly anisotropic in nature, caused by the presence of strong magnetic fields. It is useful to analyze such plasmas in a reduced-MHD limit, where fluctuation amplitudes, mode anisotropy, and fluctuation timescales are treated at linear order. In this approximation, the slow-wave and Alfvén-wave cascades are energetically decoupled in the inertial range of turbulence \citep{Schekochihin2009}. The slow- and the Alfvén-wave cascades only undergo mode mixing at length-scales comparable to the ion Larmor radius and undergo collisionless damping, channeling a fraction of the turbulent energy into heating the ions. The remaining energy cascades down to smaller length scales and dissipates at the electron Larmor radius, heating the electrons. In Figure \ref{fig:scales}, we represent the typical scales in which the above processes operate. 

In a recent paper, \citet{Kawazura2020} studied the dependence of the heat partition between the ions and the electrons on the composition of the wave cascades driving the turbulence. They established that the heating ratio between the ions and the electrons, $Q_i/Q_e$, primarily depends on the ratio of the driving powers of compressible slow waves and the incompressible Alfvén waves $P_S/P_A$, and on the plasma $\beta$. This dependence can be approximated using the expression

\begin{equation}
    \label{eq:kawazura2020}
    \dfrac{Q_i}{Q_e} = \dfrac{P_S}{P_A} + \dfrac{35}{1 + \left(\beta/15\right)^{-1.4}}.
\end{equation}

\noindent In the limit of plasma $\beta\lesssim 1$, the value of $Q_i/Q_e$ is almost entirely dependent on the composition of the driving-scale turbulent cascade, with $Q_i/Q_e\sim P_S/P_A$. Additionally, in plasmas which are dominantly driven by slow waves, nearly all of the energy carried by the slow waves is channeled into heating the ions irrespective of the value of plasma $\beta$, causing $Q_i/Q_e\sim P_S/P_A$. 

The second term in equation \eqref{eq:kawazura2020} captures the dependence of the heating ratio on the microphysics of collisionless damping of turbulence that occurs at the small lengthscales comparable to the ion gyroradius. On the other hand, the first term involving the ratio of the slow-to-Alfvén driving power, $P_S/P_A$, is determined by the large scale processes that drive turbulence at large scales. Our goal is to explore this ratio in the general setting applicable to RIAFs. 

In such flows, MRI has been identified as the most likely mechanism that drives the turbulence responsible for the transport of angular momentum~\citep{Balbus1998}. The net outward transport of angular momentum is caused by particular off-diagonal components of the Maxwell and Reynolds stress tensors. The same components of these stresses are also responsible for the injection of turbulent energy into the accretion flow \citep{Hawley1995}. To that end, we compute the rates of the injection of turbulent energy into the individual slow- and Alfvén-wave cascades, and show that we can relate them to the Maxwell and Reynolds stress tensors in a local region of the accretion flow. 

In order to infer properties of the turbulence in accretion flows, we will study the turbulence in the shearing-box approximation \citep{Goldreich1965}. This approximation allows us to capture the mechanism of the injection of turbulent energy, i.e., the MRI, in a local patch of an accretion flow. In this paper, we set up a local shearing box with mean background magnetic fields ($\vec{B_0}$) and shearing rates ($q$) relevant to black hole accretion disks, as we describe in the following section. 

\section{The Shearing Box Energy Equations in a Reduced MHD Limit}
\label{sec:Equations}

In this section, we derive the energy equations for the slow magnetosonic and the Alfvén wave fluctuations in a shearing box MHD system. We first set up the equations of MHD in a shearing box configuration. We then expand the equations to first order in the reduced-MHD approximation. In this limit, we discuss the orientations of the eigenvectors of the normal modes of MHD and derive the reduced-MHD energy equations for the Alfvén- and the slow-wave turbulent cascades. 

\subsection{The Shearing Box Approximation in the Reduced MHD Limit}

The equations of MHD in the shearing box approximation can be written as \citep{Goldreich1965}

\begin{equation}
\label{eq:continuity}
\dfrac{\partial\rho}{\partial t} + \left( \vec{u}_0 + \vec{u} \right)\cdot \vec{\nabla} \rho + \rho \left(\vec\nabla\cdot\vec u\right) = 0, 
\end{equation}

\begin{multline}
\label{eq:momentum}
\rho \dfrac{\partial \vec{u}}{\partial t} + \rho \left[\left( \vec{u}_0 + \vec{u} \right)\cdot \vec{\nabla}\right] \left(\vec u + \vec{u}_0\right) = \\ -\vec\nabla \left(P + \dfrac{B^2}{2}\right) + \left(\vec B \cdot \vec\nabla\right) \vec B - 2\rho \vec\Omega\times\vec u,
\end{multline}
and
\begin{equation}
\label{eq:induction}
\dfrac{\partial\vec B}{\partial t} = \vec\nabla\times\left[\left(\vec{u}_0 + \vec{u} \right)\times\vec B\right].
\end{equation}

\noindent In the above equations, $\rho$ denotes the plasma density, $P$ the pressure, $\vec u$ the velocity perturbation on the background shear flow $\vec{u}_0$, $\vec B$ the magnetic field strength, and $\vec\Omega$ the angular velocity of the system. We set up the equations in a Cartesian comoving coordinate system $(x,y,z)$ such that the angular velocity $\vec{\Omega} \equiv \Omega\ \hat{z}$ and $\hat{x}$ represents the radial outward orientation in a global disk. The background shear flow $\vec{u}_0$ is given by 

\begin{equation}
\label{eq:shear flow}
\vec{u}_0 = -q\Omega x \hat{y},  
\end{equation}

\noindent where $q\equiv d(\ln\Omega)/d(\ln r)$ is the shearing rate. We assume an adiabatic equation of state, $P=K\rho^{\gamma}$, where $\gamma$ is the adiabatic index and $K$ is a constant. 

\begin{figure}
\includegraphics[width =  0.47\textwidth]{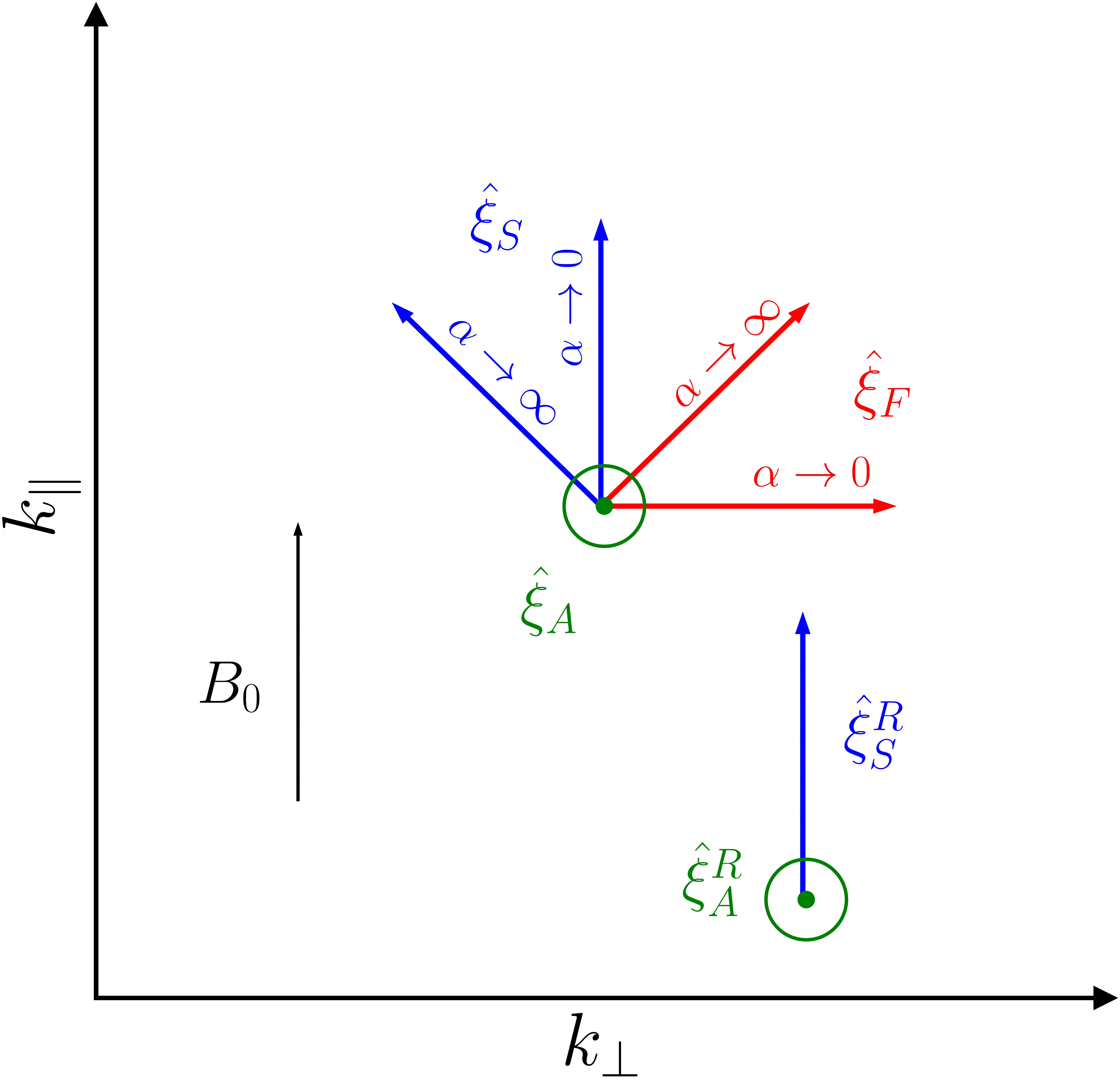}
\caption{The velocity eigenvectors for the Alfvén ($\hat{\xi}_{A}$, green), the slow magnetosonic ($\hat{\xi}_{S}$, blue), and the fast magnetosonic ($\hat{\xi}_{F}$, red) MHD waves. The green dot for the Alfvén eigenvector represents an out of plane orientation. The slow and the fast mode eigenvectors depend on the value of $\alpha$. The $\alpha \to 0$ and $\alpha \to \infty$ limits of these eigenvectors are indicated by the labels. The reduced-MHD limit of the Alfvén eigenvector ($\hat{\xi}_{A}^{R}$) is the same as the MHD eigenvector, while the same limit of the slow wave eigenvector ($\hat{\xi}_{A}^{R}$) is along the direction of the mean magnetic field (shown in the bottom-right of this figure). }
\label{fig:MHD eigenmodes}
\end{figure}

In a magnetized system described by the equations of MHD, the linear normal modes consist of the Alfvén, the slow magnetosonic, and the fast magnetosonic waves. The direction of the mean magnetic field (denoted by $\vec{B_0}$) sets the orientation of the fluctuations in each of these normal modes. The orientations of the velocity eigenvectors in Fourier space for each of the normal modes can be written as \citep[see][]{ChoLazarian2003}

\begin{equation}
\label{eq:Alfv\'en displacement eigenvector}
\hat{\xi}_{A} = -\hat{k}_{\parallel} \times \hat{k}_{\perp}
\end{equation}
\noindent for the Alfvén wave eigenvector $\hat{\xi}_{A}$, 
\begin{equation}
\label{eq:slow displacement eigenvector}
\hat{\xi}_{S} \propto (-1 + \alpha - \sqrt{D})\vec{k}_{\parallel} + (1 + \alpha - \sqrt{D})\vec{k}_{\perp} 
\end{equation}
\noindent for the slow wave eigenvector $\hat{\xi}_{S}$, and 
\begin{equation}
\label{eq:fast displacement eigenvector}
\hat{\xi}_{F} \propto (-1 + \alpha + \sqrt{D})\vec{k}_{\parallel} + (1 + \alpha + \sqrt{D})\vec{k}_{\perp} 
\end{equation}
\noindent for the fast wave eigenvector $\hat{\xi}_{F}$. In equations \eqref{eq:Alfv\'en displacement eigenvector}-\eqref{eq:fast displacement eigenvector}, the direction of the mean magnetic field sets the orientation of the parallel wavevector, with $\hat{k}_\parallel = \hat{B}_0$. The perpendicular wavenumber ($\hat{k}_{\perp}$) is defined such that $\vec{k} = \vec{k}_{\parallel} + \vec{k}_{\perp}$. The parameter $\alpha$ is defined as the square of the ratio of the sound speed ($c_s$) to the Alfvén speed ($v_A = B_0/\sqrt{4\pi\rho_0}$), i.e., 
\begin{equation}
\label{eq:alpha defn}
\alpha \equiv \dfrac{c_s^2}{v_A^2} = \dfrac{c_s^2}{B_0^2/4\pi\rho_0}.
\end{equation}
\noindent The quantity $D$ is given by $D = (1+\alpha)^2 - 4\alpha\cos^2{\theta}$, where $\cos{\theta} = |\vec{k}_{\parallel}|/|\vec{k}|$. 

Figure \ref{fig:MHD eigenmodes} shows the set of these three orthogonal eigenvectors in Fourier space. As denoted by the green dot, Alfvén waves are always azimuthally polarized with respect to the mean magnetic field (eq. [\ref{eq:Alfv\'en displacement eigenvector}]). The slow (shown in blue) and fast waves (shown in red) lie on the $k_\parallel-k_\perp$ plane with the parameter $\alpha$ determining their orientations. In a magnetically dominated plasma ($\alpha\to0$), the slow waves are polarized nearly parallel to the mean magnetic field and the fast waves are polarized nearly perpendicular to the mean magnetic field. On the other hand, in a pressure-dominated plasma where $\alpha\to\infty$, the fast waves are polarized along $\hat{k}$. Importantly, \textit{the slow waves in such a limit are nearly incompressible} and are polarized in the $-\hat\theta\equiv(\hat{k}_{\perp}\times\hat{k}_{\parallel})\times\hat{k}$ direction. They are also referred to as the pseudo-Alfvén waves \citep{Maron2001}.

The reduced MHD limit of the MHD equations is obtained by expanding the equations to first order in mode anisotropy $k_{\parallel}/k_{\perp}$ and in the perturbations of density ($\delta\rho/\rho_0$), velocity ($u/v_A$), and magnetic fields ($\delta B/B_0$) \citep[see][]{Schekochihin2009}. The frequencies $\omega$ of the fluctuations in all the fluid quantities are considered to be of the order of the frequencies of the Alfvén waves ($\omega\sim k_\parallel v_A$). As a result, the fast wave displacements exhibit timescales much shorter than the characteristic timescale $1/\omega$ and, hence, are ordered out.

We now examine the polarizations of the Alfvén- and the slow-waves in this limit by expanding the eigenvectors $\hat{\xi}_{A}$ and $\hat{\xi}_{S}$, respectively, to the leading order in $k_\parallel/k_\perp$. The polarization of the Alfvén modes in the Fourier space are identical to that in a full MHD system, $\hat{\xi}_{A}^{R} = -\hat{k}_{\parallel} \times \hat{k}_{\perp}$. The compressible slow wave components, on the other hand, undergo displacements parallel to the mean magnetic field \citep[$\hat{\xi}_{S}^{R} = \hat{k}_{\parallel}$, see eq. \ref{eq:slow displacement eigenvector} and also][]{Schekochihin2009}. Henceforth, we denote the slow wave component of the velocity and magnetic fields as $\vec{u}_\parallel$ and $\delta \vec{B}_\parallel$, respectively, and their Alfvén wave components as $\vec{u}_\perp$ and $\delta \vec{B}_\perp$. We show the orthonormal reduced MHD eigenvectors for the slow and the Alfvén waves in Figure \ref{fig:MHD eigenmodes} with the vectors labeled $\xi_S^{R}$ and $\xi_A^{R}$, respectively. \footnote{We note that, as a result of the first order expansion in anisotropy $k_\parallel/k_\perp$, the orientation of the eigenvectors are independent of the quantity $\alpha$.}

Expanding to the first order in the reduced-MHD ordering transforms equations \eqref{eq:continuity}-\eqref{eq:induction} and the equation of state to \citep[see][]{Kawazura2022} 

\begin{multline}
\label{eq:RMHD momentum}
\dfrac{\partial \vec{u}}{\partial t} + \vec{u}_{\perp}\cdot \nabla \vec{u} + \vec{u}_0\cdot\nabla \vec{u} = -\nabla_{\perp}\left(\frac{\delta B^2}{2 \rho_0}\right) \\ + \frac{1}{\rho_0}\left(\vec{B}_0\cdot\nabla\delta \vec{B} + \delta \vec{B}_\perp\cdot\nabla\delta \vec{B}\right) +  2 \vec{\Omega}\times \vec{u}\ - \vec{u}\cdot\nabla\left(\vec{u}_0\right) 
\end{multline}
and 
\begin{multline}
\label{eq:RMHD induction}
\left(\dfrac{\partial}{\partial t} + \vec{u}_\perp\cdot\nabla\right)\left(\delta\vec{B} + \dfrac{1}{\alpha}\delta\vec{B}_\parallel \right) + \vec{u}_0\cdot\nabla \left(\delta \vec{B} + \dfrac{1}{\alpha}\delta\vec{B}_\parallel\right) = \\ \delta \vec{B}\cdot\nabla\left(\vec{u}_0\right) + \left(\vec{B}_0\cdot\nabla\delta \vec{u} + \delta \vec{B}_\perp\cdot\nabla\delta \vec{u}\right).
\end{multline}
\noindent In the above equations, $\nabla_\perp$ denotes the component of the gradient operator $\nabla$ perpendicular to the mean magnetic field $\hat{B}_0$. Equations \eqref{eq:RMHD momentum} and \eqref{eq:RMHD induction} carry the following key physical manifestations resulting from the reduced MHD ordering:
\textit{(i)} the Goldreich-Sridhar critical balance assumption $k_\parallel v_A \sim k_\perp u_\perp$ leads the derivatives of the form $\vec{x}_{\parallel}\cdot\nabla \vec{x}$ to become dynamically unimportant, where $\vec{x}$ represents velocity or magnetic field fluctuations;  
\textit{(ii)} the fast magnetosonic wave fluctuations are ordered at timescales too fast to excite, resulting in a perpendicular balance between the pressure fluctuations and the magnetic tension in the direction perpendicular to the magnetic field ($\delta P = -\vec{B_0}\cdot\delta\vec{B}$); and 
\textit{(iii)} the induction equation is combined with the continuity equation to capture the compressive density fluctuations (which, in turn, is related to the parallel component of the magnetic field via the perpendicular pressure balance).

\subsection{The Energy Equations for Slow- and Alfvén-Waves}

In order to obtain the energy equation for the Alfvén polarized fluctuations, we take the the dot product of equation \eqref{eq:RMHD momentum} with $\rho_0\vec{u}_\perp$ and of \eqref{eq:RMHD induction} with $\delta\vec{B}_\perp$ and add the resulting equations. The energy equation of slow wave fluctuations can be similarly obtained by taking the the dot product of equation \eqref{eq:RMHD momentum} with $\rho_0\vec{u}_\parallel$ and of \eqref{eq:RMHD induction} with $\delta\vec{B}_\parallel$ and adding the resulting equations. We then write the energy equations at the first order in the reduced-MHD approximation as  

\begin{multline}
\label{eq:Alfv\'en Energy Equation}
\dfrac{\partial E_A}{\partial t} + \nabla \cdot \left(\vec{u}_0 \ E_A\right)  + \nabla_\parallel\cdot\left\{\rho_0\vec{v}_A \left(\vec{u}_\perp\cdot \delta \vec{B}_\perp\right)\right\} \\ + \nabla_\perp\cdot\left\{\vec{u}_\perp \ \left(E_A + \frac{\delta B^2}{2}\right) +  \delta\vec{B}_\perp \left(\vec{u}_\perp\cdot \delta \vec{B}_\perp\right)\right\} = R_A   
\end{multline} 
for the Alfvén waves and
\begin{multline}
\label{eq:Slow Energy Equation}
\dfrac{\partial E_S}{\partial t} + \nabla \cdot \left(\vec{u}_0 \ E_S\right) + \nabla_\parallel\cdot\left\{\rho_0\vec{v}_A \left(\vec{u}_\parallel\cdot \delta \vec{B}_\parallel\right)\right\} \\ + \nabla_\perp\cdot\left\{\vec{u}_\perp \ E_S +  \delta\vec{B}_\perp \left(\vec{u}_\parallel\cdot \delta \vec{B}_\parallel\right)\right\} = R_S \ \\  
\end{multline} 

\noindent for the slow waves. In these equations, we denote the total Alfvén- and slow-wave energies by $E_A$ and $E_S$, respectively, where

\begin{equation}
\label{eq:Alfv\'en Total Energy}
E_A = \frac{1}{2}\rho_0 u^2_\perp + \frac{1}{2}\delta B_\perp^2
\end{equation}
\noindent and 
\begin{equation}
\label{eq:Slow Total Energy}
E_{S} = \frac{1}{2}\rho_0 u^2_\parallel + \frac{1}{2} \delta B_\parallel^2 \left(1+\frac{1}{\alpha}\right).
\end{equation}

The right hand sides of equations \eqref{eq:Alfv\'en Energy Equation} and \eqref{eq:Slow Energy Equation} contain the energy injection rates into the Alfvén waves ($R_A$) and the slow waves ($R_S$), respectively, and are given by

\begin{equation}
\label{eq:R_A}
R_A =  -\rho_0\vec{u}_{\perp}\cdot \left(\vec{u}_\perp \cdot\nabla \vec{u}_0\right) + \delta \vec{B}_\perp\cdot \left(\delta\vec{B}_\perp \cdot\nabla \vec{u}_0\right) + \rho_0\vec{u}_{\perp}\cdot\left(2 \vec{\Omega} \times \vec{u}\right)
\end{equation}
\noindent for the Alfvén waves and 
\begin{equation}
\label{eq:R_S}
R_S = -\rho_0\vec{u}_{\parallel}\cdot \left(\vec{u}_\perp \cdot\nabla \vec{u}_0\right) + \delta \vec{B}_\parallel\cdot \left(\delta\vec{B}_\perp \cdot\nabla \vec{u}_0\right) + \rho_0\vec{u}_{\parallel}\cdot\left(2 \vec{\Omega} \times \vec{u}\right)
\end{equation}
\noindent for the slow waves. The first two terms in the energy injection rates $R_A$ and $R_S$ represent injection caused by the background shear and the third term represents the advection of energy caused by the Coriolis force. The left hand side of Equations \eqref{eq:Alfv\'en Energy Equation} and \eqref{eq:Slow Energy Equation} contains the time derivative of the respective energies and additionally, the divergence of fluxes carried by the shear velocity, the wave fluctuations, and the mean magnetic field. 

Equations \eqref{eq:R_A} and \eqref{eq:R_S} give the expressions for the energy injection rates into the slow- and the Alfvén-wave cascades, respectively, correct to the first order in the reduced-MHD approximation. Because the slow- and Alfvén-wave cascades energetically decouple at lengthscales smaller than the injection scales of the MRI, we can infer the ratio of their relative driving power using numerical shearing-box simulations that capture the physics in the injection scales of the MRI. 

In the upcoming section, we study the physics of the energy injection described in equations \eqref{eq:R_A} and \eqref{eq:R_S} and show that energy injection rates in the slow- and Alfvén- wave cascades can be related to quantities measured at large scales, namely the Maxwell and the Reynolds stress tensors in the flow. 

\section{Physics of Energy Injection into Slow and Alfvén Waves}
\label{sec:Physics}

The injection of energy and transport of angular momentum in a local region in an accretion flow is caused by the Maxwell and Reynolds stress tensors in that region. The $ij$-th component of the Maxwell stress tensor $M_{ij}$ in such a local domain of volume $V$ in the accretion flow is defined as 

\begin{equation}
\label{eq: def maxwell}
M_{ij} = -\int_V\delta B_i\ \delta B_j
\end{equation}

\noindent and the $ij$-th component of the Reynolds stress tensor $R_{ij}$ in that domain is defined as
\begin{equation}
\label{eq: def reynolds}
R_{ij} = \int_V \rho_0\ u_i\ u_j;
\end{equation}
\noindent note the minus sign in the definition of the Maxwell stress tensor. In a local shearing-box configuration, the spatial volume is chosen such that it includes the energy injection scales. 

In a shearing-box system, the background shear velocity field $\vec{u_0}$ (eq. [\ref{eq:shear flow}]) is oriented along the azimuthal direction ($\hat y$) and exhibits a gradient along $\hat x$. Consequently, the $x-y$ components of the Maxwell and Reynolds stress tensors are responsible for the net injection of energy and the outward transport of angular momentum in the local shearing domain \citep{Hawley1995,Pessah2006}. Our aim, however, is to compute the rate of energy injection into the individual slow- and Alfvén-wave cascades, and relate them to the Maxwell and Reynolds stress tensors.

The total energy injection rates into the Alfvén- and slow-wave cascades can be obtained by integrating equations \eqref{eq:Alfv\'en Energy Equation} and \eqref{eq:Slow Energy Equation}, respectively, over the spatial volume of the shearing box. Upon carrying out the integral, the terms involving the divergences vanish as they represent transport of fluxes by the background incompressible shear flow and, hence, do not contribute to energy injection. As a result, we obtain the injection powers into the individual slow- and Alfvén-wave cascades.

The slow wave injection power $P_S$ is given by 
\begin{multline}
\label{eq:Slow Injection Equation}
P_S \equiv \int_V\dfrac{d E_{S}}{dt} = \int_V q \Omega \left[\vec{u}_\parallel\cdot\left(u_x \hat{y}\right) - \delta\vec{B}_\parallel\cdot\left(\delta B_x \hat{y}\right)\right] \\- \int_V2\Omega\vec{u}_\parallel\cdot\left(u_x\hat{y}-u_y\hat{x}\right),
\end{multline}
\noindent and the Alfvén wave injection power $P_A$ is given by
\begin{multline}
\label{eq:Alfv\'en Injection Equation}
P_A \equiv \int_V\dfrac{d E_{A}}{dt} = \int_V q \Omega \left[\vec{u}_\perp\cdot\left(u_x \hat{y}\right) - \delta\vec{B}_\perp\cdot\left(\delta B_x \hat{y}\right)\right] \\- \int_V 2\Omega\vec{u}_\perp\cdot\left(u_x\hat{y}-u_y\hat{x}\right).
\end{multline}

The background shear flow causes the injection of turbulent energy into the system, with the injection being captured by the terms in the square brackets in equations \eqref{eq:Slow Injection Equation} and \eqref{eq:Alfv\'en Injection Equation}. The second term in each of equations \eqref{eq:Slow Injection Equation} and \eqref{eq:Alfv\'en Injection Equation} describes the effects of the Coriolis force in facilitating advection of energy in the plane perpendicular to the angular velocity of the shearing box (the $x-y$ plane). 

In order to express $P_S$ and $P_A$ in terms of components of the Maxwell and Reynolds stress tensors, we now invoke the polarizations of the slow- and Alfvén-wave fluctuations in shearing-box configurations relevant to black hole accretion disks into equations \eqref{eq:Slow Injection Equation} and \eqref{eq:Alfv\'en Injection Equation}. The orientation of the mean magnetic field determines the polarizations of the slow-wave and the Alfvén-wave fluctuations (see discussion surrounding Figure \ref{fig:MHD eigenmodes}). 

In the disk regions of accretion flows, the radial components of the magnetic field undergo strong shearing in the azimuthal direction. As a result, the mean magnetic field in the disk is expected to be primarily along the azimuthal direction. In addition to large scale azimuthal fields, the turbulent structure in the disk is expected to develop vertical components of the magnetic fields that are determined by the local dynamics in the flow. The MRI is driven primarily by these weak vertical components of the magnetic field. Motivated by these considerations, we setup the mean magnetic field in the shearing box to lie on the $y-z$ plane such that $\vec{B}_0 = B_{y0}\hat{y} + B_{z0}\hat{z}$. 

Using expressions for the polarizations of the slow and the Alfvén wave fluctuations (equations [\ref{eq:slow displacement eigenvector}] and [\ref{eq:Alfv\'en displacement eigenvector}]) under the reduced-MHD approximation, the injection rates $P_S$ and $P_A$ can be written as 

\begin{multline}
\label{eq:general theta slow injection}
P_S = q\Omega\left\{\left(M_{xy} + R_{xy}\right)\cos^2\phi + \left(M_{xz} + R_{xz}\right)\cos\phi\sin\phi\right\}\\ - 2\Omega\left(R_{xy}\cos^2\phi + R_{xz}\cos\phi\sin\phi\right)
\end{multline}

\noindent and

\begin{multline}
\label{eq:general theta Alfv\'en injection}
P_A = q\Omega\left[\left(M_{xy} + R_{xy}\right)\sin^2\phi - \left(M_{xz} + R_{xz}\right)\cos\phi\sin\phi\right]\\ + 2\Omega\left(R_{xy}\cos^2\phi + R_{xz}\cos\phi\sin\phi\right),
\end{multline}

\noindent respectively, where the pitch angle $\phi$ is defined such that $\tan\phi\equiv B_{z0}/B_{y0}\equiv \Gamma$.

These equations lend themselves to a straightforward interpretation. The azimuthal background shear flow causes the energy injection into the shearing box to be experienced by the components of the Alfvén and slow velocities and the magnetic field fluctuations along $\hat{y}$. The injection rates into these individual wave cascades are proportional to the Maxwell and Reynolds stress tensors with factors of $\sin\phi$ and $\cos\phi$ that capture the orientation of the wave polarizations with respect to the shear. At the same time, the effect of the Coriolis interactions is apparent on the projections of the Alfvén and slow velocity fluctuations on the $x$ and $z$ axes, captured by the Reynolds stress tensors. It is noteworthy to observe from equations \eqref{eq:general theta Alfv\'en injection} and \eqref{eq:general theta slow injection} that the total energy injection rate into the combined slow and Alfvén wave cascade, as expected, depends only on the $xy$ components of the Maxwell and Reynolds stress tensors, i.e. $P_S + P_A = q\Omega(M_{xy}+R_{xy})$. The $x-z$ components of the stresses and the effects of the Coriolis force determine individual injection powers into the slow and Alfvén wave cascade but do not contribute to the total power injected into the system.

Combining expressions for the injection powers into slow and Alfvén wave cascades, we now write the slow-to-Alfvén driving power ratio $P_S/P_A$ as 
\begin{multline}
\label{eq:general ps over pa}
\frac{P_S}{P_A} = \dfrac{q\left[\left(M_{xy} + R_{xy}\right)- \left(M_{xz} + R_{xz}\right)\Gamma^2\right] - 2\left(R_{xy} + R_{xz}\Gamma\right)}{q\left[\left(M_{xy} + R_{xy}\right)\Gamma^2 - \left(M_{xz} + R_{xz}\right)\Gamma\right] + 2\left(R_{xy} + R_{xz}\Gamma\right)}.
\end{multline}
\noindent Equation \eqref{eq:general ps over pa} represents the ratio $P_S/P_A$ in a local domain of an accretion flow, that captures the lengthscales of energy injection into the domain. This ratio depends on the off-diagonal $x-y$ and $x-z$ components of the Maxwell and Reynolds stress tensors, the local shearing rate $q$, and the orientation of the mean magnetic field in the domain with respect to the shear (which appears as the factor $\Gamma$).  

In the presence of a predominantly azimuthal mean magnetic field, the slow wave displacements are nearly along the direction of the shear ($u_\parallel\propto\xi_{S}^{R}\propto \hat{y}$) and the Alfvén wave displacements lie in the plane perpendicular to the direction of the shear velocity ($u_\perp\cdot\hat{y} = 0$). The orientation of the slow wave displacements along the shear results in them receiving all the energy injection brought into effect by the shear. The Alfvén waves do not experience direct injection of energy via the shear velocities but only experience the Coriolis interaction via the $xy$ component of the Reynolds stress tensor. Hence, in the limit of the magnetic field of the MRI-driven system to be purely azimuthal, relevant to the midplane regions of accretion disks, the slow- to Alfvén-wave energy injection ratio depends only on the ratio of the $xy$ components of the Maxwell and Reynolds stresses and can be written as 

\begin{equation}
\label{eq:ratio expression}
\left.\dfrac{P_{\rm{S}}}{P_{\rm{A}}}\right\vert_{\vec B_0\approx B_0\hat y} = \dfrac{q\left(\dfrac{M_{xy}}{R_{xy}} + 1\right) - 2}{2}.
\end{equation}

In principle, the ratio $P_S/P_A$ can be inferred locally in global simulations of black hole accretion flows using equation \eqref{eq:general ps over pa}. However, in the interest of examining the dependence of the $P_S/P_A$ on the background plasma parameters, namely the plasma $\beta$ and the shearing rate, we approach the problem by inferring $P_S/P_A$ using the Maxwell and Reynolds stress tensors using numerical simulations of the shearing-box configuration. 

\section{Numerical Computation of Slow-to-Alfvén Wave Injection Ratio}
\label{sec:Simulations}

We perform a suite of magnetohydrodynamic shearing-box simulations to study the injection of slow and Alfvén waves during the linear and non-linear turbulent stages of the growth of the MRI. In the previous sections, we derived equations for the injection rates into these turbulent cascades that are correct to first order in the reduced-MHD approximation. We now infer the value of $P_S/P_A$ at this order, using numerical MHD simulations that resolve the scales of the injection of the MRI. 

Our approach is different from the one undertaken by \citet{Kawazura2022} to compute $P_S/P_A$. These authors performed numerical shearing-box simulations by solving the dynamical equations at first order of the reduced-MHD approximation. In contrast, we solve the MHD equations without employing the reduced-MHD approximation, but only infer the value of $P_S/P_A$ using the same approximation. 

Because \citet{Kawazura2020} solve the reduced-MHD equations, the energy cascades of the slow and the Alfvén waves are identically decoupled in their setup past the injection scales. In our setup, there can potentially be some coupling between the slow and the Alfvén wave cascades that are of higher order in the reduced-MHD approximation. Such higher order terms might also cause non-local exchanges in energy in the Fourier space between the velocity and magnetic field cascades \citep{Lesur2011}. These effects on the slow- and Alfvén-wave driving powers in MRI-driven MHD turbulence have not been explored and will be the subject of a later study.

\subsection{Numerical Setup}
We carry out local MHD simulations of unstratified shearing boxes using the Godunov-type finite volume code \texttt{Athena++} \citep{Stone2020}. We set up the box to be periodic in the $y$ (azimuthal) and $z$ (vertical) directions, and shear-periodic in the $x$ (radial) direction \citep[see][for a discussion on the shear-periodic boundary conditions]{Hawley1995}. The box dimensions are set to a size $(L_x,\ L_y,\ L_z)$ = $(4L,\ 4L,\ L)$, where $L = c_s/\Omega_0$. The resolutions of the box in the $x,\ y,\ \text{and}\ z$ directions are equal, with the number of grid points being $(N_x,\ N_y,\ N_z)$ = $(4N,\ 4N,\ N)$. We perform our simulations with $N=64$. In our simulation set, time-integrations are carried out using the second order van Leer predictor-corrector methods and spatial reconstructions are performed using piecewise linear interpolation.

\begin{table}
\caption{Background conditions explored in shearing-box simulations}
\begin{center}
\begin{threeparttable}
\noindent \begin{tabular}{ c c }
\toprule
Parameter  & Values \\
\midrule
Shearing Rate ($q$) & 0.5, 0.75, 1.0, 1.25, 1.5, 1.75 \\
Vertical $\beta$ ($\beta_z$) \tnote{a} & 800 \\
Azimuthal $\beta$ ($\beta_y$) & 2, 5, 10, 20 \\
Equation of State & Isothermal \\
\bottomrule
\end{tabular}
\begin{tablenotes}
\item[a] We define the vertical and azimuthal plasma $\beta$'s using the respective components of the mean magnetic fields over the volume of the shearing box. We set the radial component of the mean magnetic field to zero, resulting in the mean magnetic vertical and azimuthal fields to be constant in time. 
\end{tablenotes}
\end{threeparttable}
\end{center}
\label{tab:simulation library}
\end{table}

The MRI is primarily an instability driven by weak vertical magnetic fields in an accretion disk. The instability amplifies velocity and magnetic field perturbations in the radial direction as vertical modes. However, the direction of the mean magnetic field in the accretion disk midplane is expected to be predominantly azimuthal, as a result of the global scale shearing caused by the background azimuthal velocity field on radial magnetic fields. The strong azimuthal fields have the potential to become dynamically important, as they cause the growth of large-scale azimuthal modes \footnote{The effects of the curvature of magnetic field lines can be dynamically important to the growth of the MRI at strong azimuthal field strengths \citep{Pessah2005}. We do not consider these effects in our study.}. These considerations motivate us to set up the shearing box in our simulations with a strong magnetic field in the azimuthal direction combined with a weak magnetic field in the vertical direction. 

The midplane regions of accretion disks typically exhibit shearing rates equal to the Keplerian value ($q=1.5$). However, in regions away from the midplane, non-Keplerian shearing rates are expected due to non-trivial effects of pressure support arising in thick accretion flows. Additionally, in the inner regions of the flow, general relativistic corrections to the azimuthal flow profile can result in non-Keplerian shearing rates. In the interest of allowing our study to be broadly applicable to global-scale studies of accretion flows, we study the local shearing box simulations for a wide range of shearing rates. We also set the equation of state in the system to be isothermal. In Table \ref{tab:simulation library}, we list the parameter space of the background conditions of the shearing box explored in this paper.

\subsection{Linear MRI Growth and Turbulence}

\begin{figure}[t]
\includegraphics[width =  0.47\textwidth]{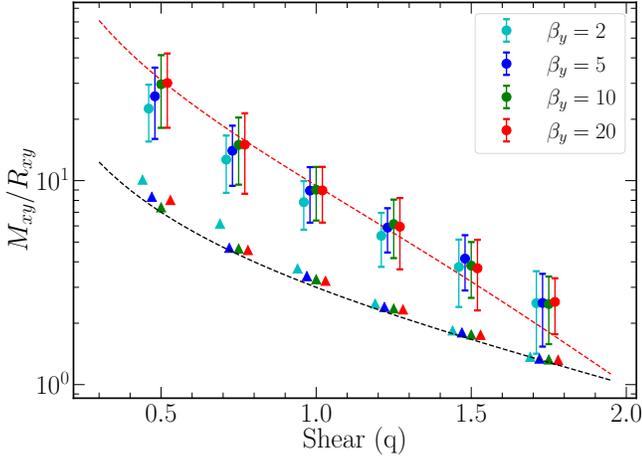}
\caption{The ratio of the volume-integrated $xy$ component of the Maxwell stress tensor to that of the Reynolds stress tensor, plotted as a function of shearing rate, for different values of $\beta_y$, computed from our simulation set. The triangles indicate the stress ratios at the stage of linear growth of the MRI (channel solutions), while the filled circles represent the same ratio computed during a saturated turbulent stage at later times in the simulation ($t = 230-300 \Omega^{-1}$). The black dotted line represents the analytical calculation \citep{Pessah2006a} of the dependence of $M_{xy}/R_{xy}$ given by equation \eqref{eq:Pessah M over R general k} on the shearing rate $q$, for the fastest growing vertical mode $k_{\rm{M}} = k_{\rm{max}} = \sqrt{q(4-q)}/2$. The red dotted line represents the same analytically obtained ratio, but instead assuming that the ratio in the non-linear turbulent stage is captured by an effective wavenumber $k_{\rm{eff}} = 0.42 k_{\rm{max}}$.}
\label{fig:stress ratios}
\end{figure}

The linear growth of the MRI in the presence of a weak, vertical mean magnetic field is well understood both analytically and numerically \citep{Balbus1991,Balbus1998}. In a numerical simulation, the linear stage of growth is dominated by the dynamics of the fastest growing mode of the MRI. The simulations at this stage take the form of channel solutions where the velocity and the magnetic fluctuations in the $xy$ plane undergo exponential growth with a rate equal to the growth rate of the fastest growing mode. These channel solutions are vertical, axisymmetric modes, and also satisfy the complete non-linear equations of motion. At later times of their exponential growth, the structure of the channel solutions makes them unstable to Kelvin-Helmholtz-like parasitic instabilities, which breaks the solutions into a non-axisymmetric turbulent state \citep{GoodmanXu1994,Pessah2010}. In the presence of strong azimuthal magnetic fields, large wavelength non-axisymmetric modes are also unstable to the MRI.

The outward transport of angular momentum is brought into effect by the $xy$ components of the Maxwell and Reynolds stress tensors. \citet{Pessah2006a} showed that the ratio of the volume-integrated Maxwell to Reynolds stress tensor at late times of linear MRI growth in the presence of a weak vertical background magnetic field depends on the wavenumber $k_{\rm{M}}$ of an unstable mode and the shearing rate ($q$), and is given by 

\begin{equation}
\label{eq:Pessah M over R general k}
\dfrac{M_{xy}(k_{\rm{M}})}{R_{xy}(k_{\rm{M}})} = 1 + \dfrac{2(2-q)}{k_{\rm{M}}^2 + \gamma(k_{\rm{M}})^2}.
\end{equation} 

\noindent In equation \eqref{eq:Pessah M over R general k}, the quantity $\gamma(k_{\rm{M}})$ represents the growth rate of the unstable mode with wavenumber $k_{\rm{M}}$, and is given by

\begin{equation}
\label{eq:Pessah M over R general k}
\gamma(k_{\rm{M}}) \equiv \left\{\sqrt{(2-q)^2 + k_{\rm{M}}^2}-\left[k_{\rm{M}}^2 + (q-2)\right]\right\}^{1/2}.
\end{equation} 

\noindent Assuming that the growth of the linear MRI is dominated by the fastest growing mode ($k_{\rm{M}}=k_{\rm{max}} = \sqrt{q(4-q)}/2$), the ratio of the Maxwell to the Reynolds stress tensor in this stage can be approximated as 

\begin{equation}
\label{eq:Pessah M over R}
\left.\dfrac{M_{xy}}{R_{xy}}\right\vert_{\text{linear MRI}} \approx \dfrac{4-q}{q}.
\end{equation}

In our set of numerical simulations, we observe a period of channel-solution-like growth of the linear MRI during the early stages of each simulation. During these periods, we numerically calculate the ratio $M_{xy}/R_{xy}$. In Figure \ref{fig:stress ratios}, we plot this ratio as a function of the shearing rate using triangular dots. We observe that the ratio is nearly independent of the plasma $\beta$ but primarily depends on the shearing rate $q$. Additionally, we see that the ratio closely follows the analytical result for $M_{xy}/R_{xy}$ for the fastest growing mode given by equation \eqref{eq:Pessah M over R}. These observations establish that the signature of the fastest growing vertical mode on the volume integrated stress ratio $M_{xy}/R_{xy}$ is persistent even in the presence of significant azimuthal background magnetic fields. 

In order to study the non-linear turbulent stage, we compute the ratio $M_{xy}/R_{xy}$ at later times of our simulations ($t = 230-300 \Omega^{-1}$) \footnote{In order to compute the Reynolds stress tensor, we use the local fluid density $\rho$ as opposed to the background density $\rho_0$ (see eqn. [\ref{eq: def reynolds}]), which is consistent with the order of the calculation.}. At this stage, the ratio attains values higher than the corresponding ratio at the the linear stage of growth. This behavior is observed for all values of shearing rate, independent of the background plasma $\beta$. The numerically computed ratios are shown as the circular dots in Figure \ref{fig:stress ratios}, with the overplotted error bars showing the measured standard deviation in the ratios. While this behavior is known in shearing box systems in the presence of a weak vertical mean magnetic fields \citep[see Figure 3 in][]{Pessah2006a}, the addition of strong azimuthal magnetic fields does not seem to alter their conclusion. 

We now aim to provide a simplistic model for the shear-dependence of the ratio $M_{xy}/R_{xy}$ during the non-linear turbulent stage. We posit that at the late times, the dynamics of the turbulence can be captured by an effective MRI-unstable mode. We model this effective mode such that its wavenumber $k_{\rm{eff}} = a k_{\rm{max}}$, where $a$ is a scaling parameter and $k_{\rm{max}}=\sqrt{q(4-q)}/2$ is the wavenumber of the fastest growing linear mode. In Figure \ref{fig:stress ratios}, we plot the theoretical shear dependence of the ratio $M_{xy}/R_{xy}$ for a wavenumber $k_{\rm{eff}}$ such that $a=0.42$, demonstrating that the dependence of $M_{xy}/R_{xy}$ on the shearing rate is well captured by this choice.

While the contribution of the azimuthal component of the mean magnetic field is subdominant in the dynamics of the MRI, it does have important implications on the injection of energy into the slow and the Alfvén wave cascades. First, as discussed in \S \ref{sec:Equations}, the large-scale mean field in the flow determines the polarizations for the slow- and the Alfvén-wave displacements. The relative orientations of the polarizations of modes and the direction of the shear is key to determining the energy injection rates into the individual cascades. Additionally, the persistent large-scale mean field in the disk causes the cascade of turbulence to be anisotropic. The strong anisotropy in the turbulence results in the decoupling of the wave cascades past the injection scales. In the following subsection, we infer the slow- to Alfvén-wave driving power from the Maxwell and Reynolds stress tensors computed here.

\subsection{Slow- to Alfvén-Wave Driving Power Ratio}

\begin{figure}[t]
\includegraphics[width =  0.47\textwidth]{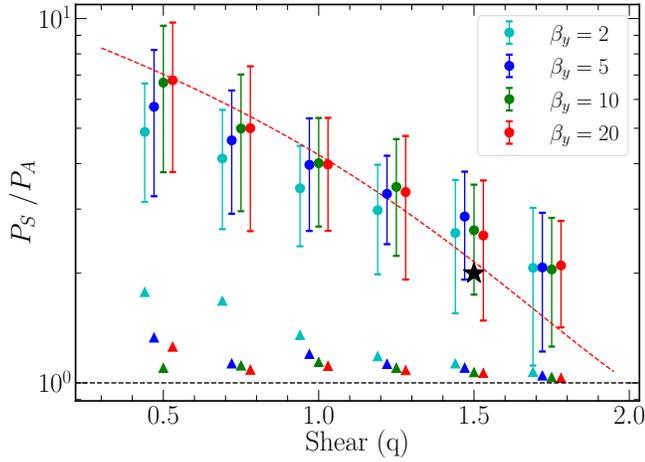}
\caption{The ratio of slow wave injection rate to the Alfvén wave injection rate, $P_S/P_A$, plotted as a function of shearing rate, computed for our simulation set. The triangular dots indicate the inferred ratio at the stage of linear growth of the vertical mode MRI (channel solutions), and the circled dots represent the same ratio computed during a saturated turbulent stage at later times in the simulation ($t=230-300 \Omega^{-1}$). The dotted line represents the value of $P_S/P_A$ calculated analytically for the linear stage using Equation \eqref{eq:linear injection ratio}. The black star indicates the calculation of $P_S/P_A$ by \citet{Kawazura2022} who solved the dynamical equations of MHD at the first order in the reduced-MHD approximation. }
\label{fig:heating ratios}
\end{figure}

Having numerically computed the volume integrated Maxwell and Reynolds stress tensors in MRI driven turbulence with predominantly azimuthal mean fields, we now translate them into the slow-to-Alfvén injection power ratio using equation \eqref{eq:ratio expression}. In Figure \ref{fig:heating ratios}, we plot the inferred $P_S/P_A$ as a function of shearing rate for all the background magnetic field strengths explored in our set of simulations. 

The ratio $P_S/P_A$ assumes a value of near unity during the linear stage of MRI growth, for almost all the our simulations spanning various shearing rates and values of plasma $\beta$. This result can be analytically explained by combining the theoretical value of $M_{xy}/R_{xy}$ for the fastest growing vertical mode (eq. [\ref{eq:Pessah M over R}]) with equation \eqref{eq:ratio expression} as 

\begin{equation}
\label{eq:linear injection ratio}
\left.\dfrac{P_{\rm{S}}}{P_{\rm{A}}}\right\vert_{\text{linear MRI},\ \vec B_0\approx B_0\hat y} = \dfrac{q\left(\left.\dfrac{M_{xy}}{R_{xy}}\right\vert_{\text{linear MRI}} + 1\right) - 2}{2} = 1.
\end{equation}

\noindent Remarkably, even though the ratio $M_{xy}/R_{xy}$ during the linear growth of the MRI carries a strong dependence on the shearing rate $q$ (eq. [\ref{eq:Pessah M over R}]), the value of $P_S/P_A$ during this stage is independent of the shearing rate.  

In the non-linear stage of the MRI, the ratio $P_S/P_A$ takes values higher than in the linear stage, with a weak dependence on shear. The values of $P_S/P_A$ range between $\sim 2-8$ depending on the shearing rate. The effects of non-linear interactions in the turbulence cause the ratio $M_{xy}/R_{xy}$ to be significantly higher than the corresponding linear limit at low values of $q$. Hence, the ratios $P_S/P_A$ are significantly higher at lower values of shearing rates. 

The background plasma $\beta$ in the shearing box does not have a strong effect on the ratio $P_S/P_A$ at a given value of a shearing rate. This is a result of the fact that the signature of MRI-driven turbulence on the ratio of Maxwell and Reynolds stress tensors, even in the non-linear stages of growth, follows the trend determined by the fastest growing linear modes (Figure \ref{fig:stress ratios}). As discussed in the previous subsection, the ratio of the stress tensor depends only on shear and is independent of the background magnetic field strength. 

For shearing rates equal to a Keplerian value, $q=3/2$, which is expected in the midplanes of black-hole accretion flows, the value of $P_S/P_A$ is approximately 2. This computation is in agreement with the calculation of the same quantity performed using reduced-MHD simulations by \citet{Kawazura2022}, which we show as a black star in Figure \ref{fig:heating ratios}. The ratio computed by \citet{Kawazura2022} is also independent of plasma $\beta$, in agreement with our findings and physical explanation as described in \S \ref{sec:Physics}. 

While our calculation of $P_S/P_A$ is a first order estimate of the value under the reduced-MHD approximation computed from MHD simulations, the agreement with \citet{Kawazura2022} indicates that the approximation is largely valid in the parameter regime that we consider. From our analysis, it is evident that irrespective of the background plasma $\beta$ and the shearing rate in an MRI driven turbulent flow, the fraction of turbulent energy injected into the slow waves is significantly higher than the energy injected into the Alfvén waves.

\subsection{A Sub-grid Model for the Slow-to-Alfvén Driving Power and Ion-to-Electron Heating Ratio}

\begin{figure}[t]
\includegraphics[width =  0.47\textwidth]{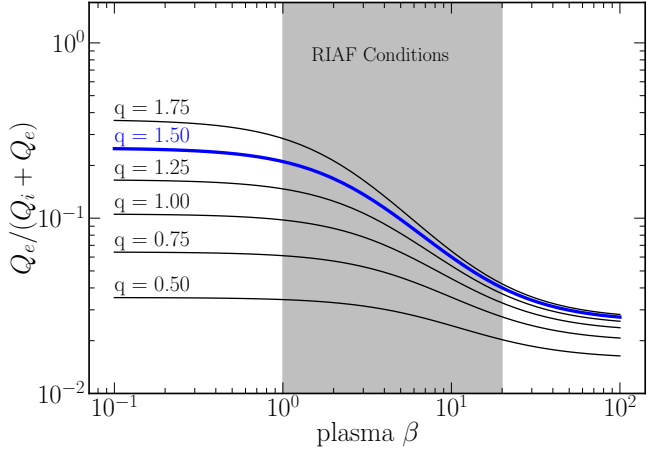}
\caption{The electron heating fraction, $Q_{\rm{e}}/\left(Q_{\rm{i}} + Q_{\rm{e}}\right)$, inferred from eq. \eqref{eq:heating ratio fit} as a function of plasma $\beta$, plotted for different values of the shearing rate ($q$). The electron heating fraction dependence on plasma $\beta$ corresponding to a Keplerian shearing rate is highlighted in blue. The dashed line depicts the electron heating fraction at the limit in which the turbulence is purely driven by Alfvén waves. The typical values of plasma $\beta$ encountered in GRMHD simulations of accretion flows are highlighted in the shaded region. }
\label{fig:heating ratios total}
\end{figure}

In the discussion surrounding Figure \ref{fig:stress ratios}, we suggested that the value of $M_{xy}/R_{xy}$ during the non-linear stage of the turbulence can be captured by the theoretical estimate of the ratio for an effective wavenumber $k_{\rm{eff}}$. We demonstrated that an approximate value for $k_{\rm{eff}} = 0.42 k_{\rm{max}}$, where $k_{\rm{max}}$ is the wavenumber of the fastest growing vertical MRI mode. Using equation \eqref{eq:ratio expression}, we then write    

\begin{equation}
\label{eq:ratio fit}
\left.\dfrac{P_{\rm{S}}}{P_{\rm{A}}}\right\vert_{\vec B_0\approx B_0\hat y} = \dfrac{1}{2}\left\{q\left[2 + \dfrac{2(2-q)}{k_{\rm{eff}}^2 + \gamma(k_{\rm{eff}})^2} \right] - 2\right\}.
\end{equation}

\noindent In equation \eqref{eq:ratio fit}, the value of $k_{\rm{eff}}$ is given by 
\begin{equation}
\label{eq:k_eff}
k_{\rm{eff}} = 0.42 k_{\rm{max}} = 0.42\frac{\sqrt{q(4-q)}}{2},
\end{equation}
\noindent and the corresponding growth rate $\gamma(k_{\rm{eff}})$ for the mode $k_{\rm{eff}}$ is given by
\begin{equation}
\label{eq:k_eff}
\gamma(k_{\rm{eff}}) = \left[\sqrt{(2-q)^2 + k_{\rm{eff}}^2}-\left\{k_{\rm{eff}}^2 + (q-2)\right\}\right]^{1/2}.
\end{equation}

In the dotted red lines shown in Figure \ref{fig:heating ratios}, we depict the ratio $P_S/P_A$ given by equation \eqref{eq:ratio fit}. Clearly, this expression captures the dependence of the numerically inferred values of $P_S/P_A$ on the shearing rate $q$. We now combine this with microphysical studies by \citet{Kawazura2020} for the ion-to-electron heating ratio $Q_i/Q_e$ resultant from collisionless damping of slow- and Alfvén-wave turbulence to write (see eq. [\ref{eq:kawazura2020}]) as 

\begin{equation}
\label{eq:heating ratio fit}
\dfrac{Q_i}{Q_e} = \dfrac{1}{2}\left\{q\left[2 + \dfrac{2(2-q)}{k_{\rm{eff}}^2 + \gamma(k_{\rm{eff}})^2} \right] - 2\right\} + \dfrac{35}{1 + (\beta/15)^{-1.4}}.
\end{equation}

\noindent The plasma $\beta$-dependent second term captures the effects of mode mixing on the heating ratio at microphysical scales comparable to the ion gyroradius. The first term represents the impact of the large scale turbulent driving on the heating ratio. Notably, this impact of the large-scale driving is independent of plasma $\beta$ but primarily depends on the shearing rate $q$. 

In Figure \ref{fig:heating ratios total}, we plot the electron heating fraction, $Q_{\rm{e}}/\left(Q_{\rm{i}} + Q_{\rm{e}}\right)$, as a function of plasma $\beta$ for different values of the shearing rate $q$, using equation \eqref{eq:heating ratio fit}. For comparison, we plot the same fraction for a purely Alfvénically driven turbulence. In the shaded region, we depict the values of plasma $\beta$ that black hole accretion disks typically exhibit in GRMHD simulations. Clearly, the shearing rate-dependent nature of turbulent driving strongly determines the electron heating ratio in the disk regions of accretion flows.

\begin{figure}[t]
\includegraphics[width =  0.47\textwidth]{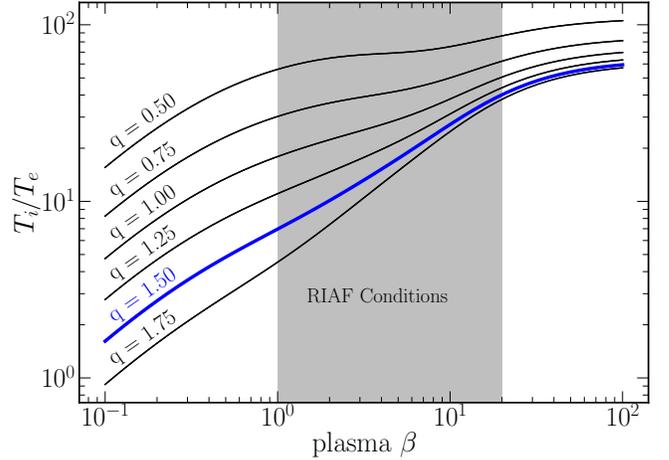}
\caption{The temperature ratio, $T_i/T_e$, as a function of plasma $\beta$, in the inner region of a radiatively inefficient accretion flow (at $\sim 10$ gravitational radii), for different values of the shearing rate, calculated using the two-temperature, covariant, semi-analytic model of \citet{Satapathy2023} and the ratio $P_S/P_A$ given by equation \eqref{eq:ratio fit}. For typical values of the plasma $\beta$ parameter and for a Keplerian shear ($q=1.5$), we find $T_i/T_e \sim 5-40$, with the ratio increasing to $\sim 80$ as the shear becomes shallower.}
\label{fig:temp ratio plot}
\end{figure}

The sub-grid model for the ion-to-electron heating ratio given by equation \eqref{eq:heating ratio fit} can be integrated with two-temperature GRMHD simulations in order to model the electron thermodynamics in radiatively inefficient accretion flows. It can also be used to infer electron temperatures from one-temperature GRMHD simulations in a way which is self-consistent with the large-scale turbulent driving. \citet{Satapathy2023} used a semi-analytic, height- and azimuthally-averaged two-temperature model for an accretion flow to calculate the ion-to-electron temperature ratio ($T_i/T_e$) taking into account the heat partition due to the slow- and Alfven-wave turbulent driving as well as compressional heating effects (see, eq.~[37] in that paper). We use this calculation combined with the shear-dependent model $Q_i/Q_e$ presented above to estimate the ion-to-electron temperature ratio in the inner accretion flow around a black hole and plot it in Figure \ref{fig:temp ratio plot} as a function of plasma $\beta$, for different values of the shearing rate. For this plot, we have set the radial distance to 10 gravitational radii and the radial profile of the radially infalling velocity to follow a power law with an index -1.5, which was obtained via calibration with GRMHD simulations by \cite{Ozel2022}. For typical values of the plasma $\beta$ parameter and for a Keplerian shear ($q=1.5$), we find $T_i/T_e \sim 5-40$, with the ratio increasing to $\sim 80$ as the shear becomes shallower.

The primary underlying assumption in the model for $Q_i/Q_e$ given by equation \eqref{eq:heating ratio fit} is that the mean magnetic field is nearly in the azimuthal direction in the region of interest in the accretion flow. This is a reasonable assumption for the disk regions of accretion flows. In certain configurations of global simulations, however, the disk evolves into a magnetically arrested state \citep[MAD, see][]{Narayan2012}. In such a state, significant vertical components of magnetic field configuration might be observed in the inner regions of the disk. Additionally, the $x-z$ components of stress tensors, which are dynamically unimportant in a local shearing box, can potentially impact the partition of energy into slow and Alfvén wave cascades due to the effects of vertical stratification. The impacts of these effects on driving the slow and Alfvén waves need to be further explored.

\section{Discussion}
\label{sec:Discussion}

In this paper, we derived an analytical expression for the ratio of the slow-wave to Alfvén-wave driving powers ($P_S/P_A$) in an MRI-driven turbulence. Working in the linear order limit of the reduced-MHD approximation, we established the connection of the ratio $P_S/P_A$ to the macroscopic properties of the turbulence. We show that the ratio depends on the components of the volume integrated Maxwell and Reynolds stress tensors in the turbulence that are responsible for the outward transport of angular momentum, and on the orientation of the mean magnetic field with respect to the direction of shear. We argue that the underlying physics that determines this ratio is the impact of shear injection of energy by the MRI and the impact of Coriolis force on the projection of the slow- and the Alfvén-wave displacements with respect to the orientation of the shear.

Using numerical shearing box simulations with magnetic field configurations relevant to black hole accretion disks, we compute the ratio $P_S/P_A$ during the non-linear saturated turbulent state of MRI-driven turbulence. During this stage, we find the ratio between the volume-integrated Maxwell and Reynolds stress tensors to be nearly independent of the background magnetic field strength. As a result, the ratio of the driving power of the compressive slow-wave cascade to that of the Alfvén-wave cascade is nearly independent of plasma $\beta$. The only dependence seen in $P_S/P_A$ is a relatively weak dependence on the shearing rate. We combine our inference of a shear-dependent model for $P_S/P_A$ with hybrid kinetic-scale studies of ion and electron heating \citep{Kawazura2020} to derive a local sub-grid model for the ion-to-electron heating ratio in black hole accretion flows.

Two-temperature GRMHD simulations that evolve the global thermodynamics of the ions and the electrons allow for the direct study of the impact of plasma heating \citep{Ressler2015,Sadowski2017}. Recent studies, such as the ones performed by \citet{Chael2018a} and \citet{Dexter2020} explore the impact of the electron heating physics on the global thermodynamics in GRMHD simulations and their implications on observational signatures. These studies show that the subgrid models for particle heating have a significant impact on not only the dynamics of the accretion flows but also on their implied images and spectra. However, in all of the above studies, subgrid heating of the plasma species due to collisionless damping has been performed assuming pure Alfvén-wave cascades in the turbulence. Our study offers a way to accurately model the ion-to-electron heating ratio considering the injection of energy into slow and Alfvén waves locally in an accretion flow based on components of the Maxwell and Reynolds stress tensors. 

A number of additional dissipation mechanisms can effect the partition of heat between the ions and the electrons. Such mechanisms include episodes of magnetic reconnection \citep{Ball2018, Ball2019} and dissipation of the stress resulting from anisotropies in velocity space \citep{Sharma2007,Kunz2018}. The effects of these mechanisms and their connections to global quantities in accretion flows need to be further explored. 

Using a height- and azimuthally-averaged covariant two-temperature model for an accretion disk, \citet{Satapathy2023} showed that the electron temperature in the inner regions of the disk is determined primarily by the slow waves in the turbulence. Incorporating local stress-tensor dependent models of sub-grid heating in accretion flows will make the small-scale dissipation processes self consistent with the large scale physics of energy injection.

\acknowledgements
The authors acknowledge support from NASA ATP award 80NSSC20K0521 and NSF PIRE award OISE-1743747. We thank K.G. Klein, M. Golden, M. Avara, T. Trent, and K. Roley for useful discussions. 

\appendix
\section{Relation of Slow, Fast, and Alfvén Modes to the Helmholtz Decomposition}
\label{Appendix A}

As discussed in the main text, the partition of turbulent heat between ions and electrons was calculated by \citet{Kawazura2020} using hybrid gyrokinetic simulations, in the presence of what these authors called ``compressive-'' and Alfv\'en-wave cascades. \citet{Kawazura2022} then performed reduced-MHD simulations in a shearing box to  compute numerically the compressive-to-Alfvénic driving power ratio and found it to be $\simeq 2$. These results have been questioned recently by two parallel studies: \citet{Zhdankin2021} used relativistic particle-in-cell simulations and found only qualitative similarities to the results of \citet{Kawazura2020}; \citet{Bacchini2024} have attempted to compute the compressive-to-Alfv\'en ratio from kinetic shearing-box simulations of a pair plasma and found it to be significantly smaller than unity, contradicting the results of \citet{Kawazura2022}. 

In this Appendix, we show that, even though both sets of studies discuss ``compressive'' and Alfv\'enic cascades, they actually decompose the turbulent cascades into different sets of modes. As a result, the only difference between the two sets of studies appears to be their definitions of the nature of the ``compressive'' cascade. \citet{Zhdankin2021} and \citet{Bacchini2024} defined compressive fluctuations in the sense of the Helmholtz decomposition, i.e., following the literal meaning of the word compressive and distinguishing modes based on their velocity divergence.  Consequently, their compressive-wave cascade consists primarily of fast wave fluctuations, which, as noted by the authors, are expected to result from spiral shock waves in the large-scale system. However, fast magnetosonic waves are ordered out in the reduced-MHD calculation of \citet{Kawazura2022} so that their label of ``compressive'' cascade consists purely of slow magnetosonic fluctuations.  It is this ratio of this slow magnetosonic cascade to the Alfv\'enic one that determines the partition of heat between ions and electrons.

In this paper, we have followed the decomposition of the field into slow, fast, and Alfvén modes derived by \citet[][]{ChoLazarian2003}. We use this to demonstrate the reason why the Helmholtz decomposition is not useful to separate slow and Alfvén modes in an MHD system, and to re-iterate that the nature of turbulent driving should be measured by the power in the individual slow, fast, and Alfvén wave cascades. 

\begin{figure}
    \centering
    \includegraphics[width =  0.47\textwidth]{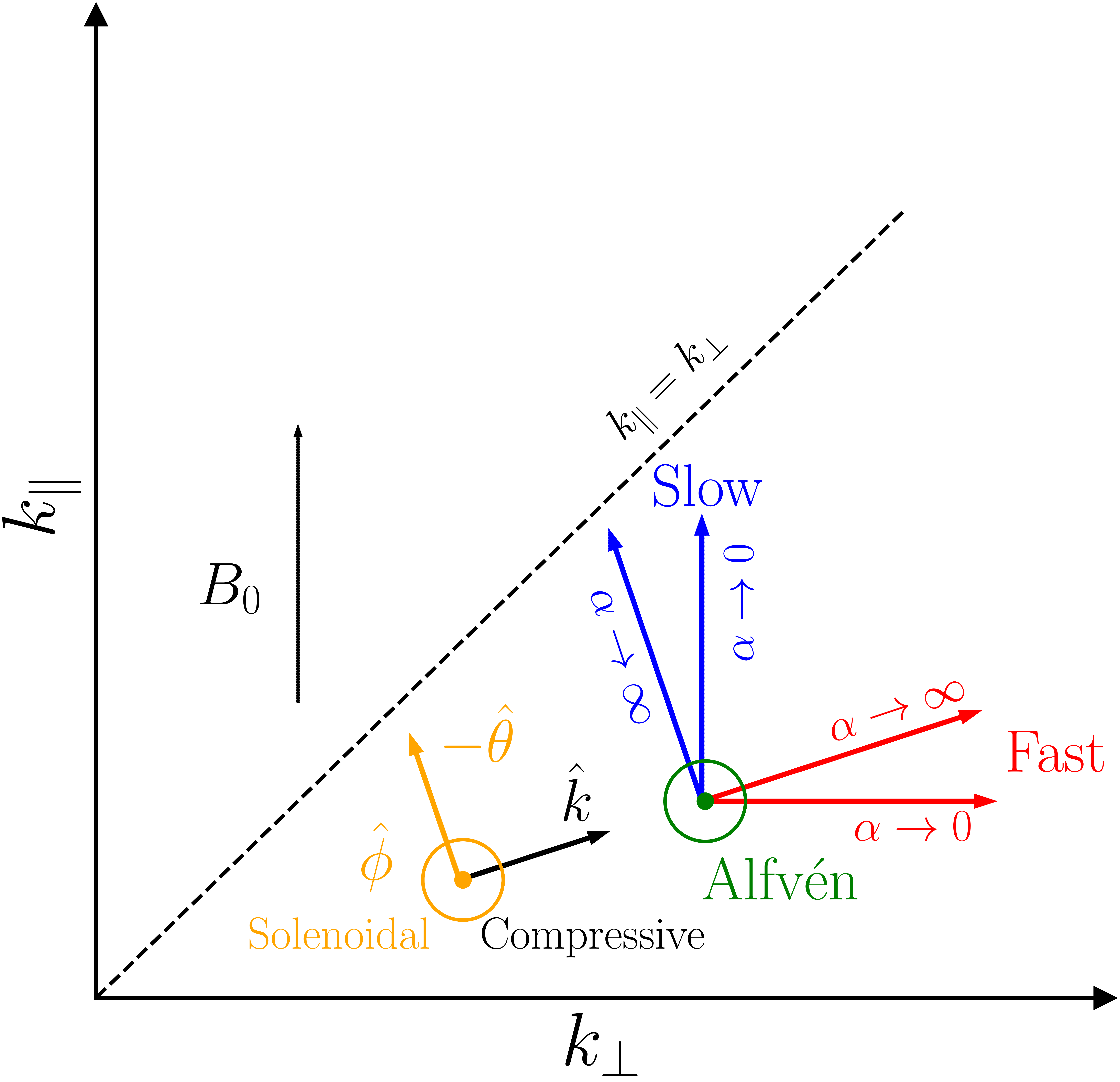}
    \caption{A representative plot to illustrate the difference between the decomposition of the plasma velocity field into the slow, fast and Alfvén modes, and the Helmholtz decomposition into compressive and solenoidal modes.  The slow and fast wave velocity eigenvectors are shown in the blue and red arrows respectively. Their orientations are dependent on the parameter $\alpha\equiv c_s^2/v_A^2$, where $c_s$ is the sound speed and $v_A$ is the Alfvén speed. The Alfvén wave eigenvector is shown in the green circled dot, and is independent of $\alpha$. The basis for the Helmholtz decomposition is shown for a equal value of $k_\parallel/k_\perp$. The solenoidal components of a vector field resulting from the Helmholtz decomposition lie in the $\hat\theta-\hat\phi$ plane indicated in orange, while the compressive components lie along $\hat k$, shown in black. }  
    \label{fig:helmholtz vs mhd}
\end{figure}




The Helmholtz decomposition of a velocity field in Fourier space $\vec{\Tilde{v}}(\vec{k})$ can be written as 

\begin{equation}
    \label{eq:helmholtz}
    \vec{\Tilde{v}}(\vec{k}) = \dfrac{\vec{k}\cdot\vec{\Tilde{v}}(\vec{k})}{\vec{k}\cdot\vec{k}} \vec{k} + \dfrac{\vec{k}\times\{\vec{\Tilde{v}}(\vec{k})\times\vec{k}\}}{\vec{k}\cdot\vec{k}}.
\end{equation}

\noindent In the right hand side of equation \eqref{eq:helmholtz}, the first and the second terms represent the curl-free and the divergence-free components of the vector field respectively. If the vector field represents a plasma velocity the curl-free component represents the compressible part of the flow, and the divergence-free component represents the incompressible part of the flow. 

In Figure \ref{fig:helmholtz vs mhd}, we use the black arrows to represent the basis of the Helmholtz decomposition. At a given point in the Fourier space defined by the parallel ($k_\parallel$) and perpendicular ($k_\perp$) wavenumbers, the first term in equation \eqref{eq:helmholtz} picks out components of the velocity field along $\hat{k}$. The second term on the other hand, represents components of the velocity field that lie on the plane perpendicular to $\hat{k}$, i.e. the $\hat{\theta}-\hat{\phi}$ plane. 

We now re-examine the eigenvectors of the linear modes of MHD, namely the slow, fast, and Alfvén waves, that we introduced in \S \ref{sec:Equations}. Following the notation of Figure \ref{fig:helmholtz vs mhd}, the orientations of the slow and the fast wave eigenvectors are represented by the blue and red arrows respectively. These eigenvectors and the basis defined by the Helmholtz decomposition are represented at points in the Fourier space such that they have the same $k_\parallel/k_\perp = 1/3$. Depending upon the value of $\alpha$ (equation \ref{eq:alpha defn}), the slow waves are oriented between $\hat{k}_\parallel$ and $-\hat{\theta}$, and the fast waves are oriented between $\hat{k}_\perp$ and $\hat{k}$. The Alfvén wave eigenvector orientation, shown in the green dot in Figure \ref{fig:helmholtz vs mhd}, is independent of $\alpha$ and is always in the $\hat{\phi}$ direction. 

We now turn to compare the eigenvectors of the linear MHD modes with the basis defined by the Helmholtz decomposition. It is to be noted that the Helmholtz decomposition only picks out the components of a velocity field along and perpendicular to $\hat{k}$ (shown by the black arrow in Figure \ref{fig:helmholtz vs mhd}), and define a solenoidal basis in the plane perpendicular to $\hat{k}$ (shown in orange). The Alfvén waves, being polarized along $\hat{\phi}$, however, always lie in the plane perpendicular to $\hat{k}$. This is consistent with the physical picture of Alfvén waves being incompressible modes. 

On the other hand, both the slow and fast wave eigenvectors in general have components of velocity along $\hat{k}$ and in the plane perpendicular to $\hat{k}$. In the limit of $\alpha\to\infty$, the slow waves lie entirely in the plane perpendicular to $\hat{k}$ and the fast waves are polarized along $\hat{k}$. These properties of the orientations of the linear MHD modes clarifies that the Helmholtz decomposition in general cannot distinguish the Alfvén modes from the components of the slow and the fast waves perpendicular to $\hat{k}$. Particularly, in anisotropic plasmas where most of the power lies in the region $k_\parallel \ll k_\perp$, the fast mode carries most of the compressible power, in the $\hat{k}$ direction. The slow modes are nearly incompressible, with their polarisation nearly along $-\hat{\theta}$ and for this reason, they are also referred to as pseudo-Alfvén modes \citep{Maron2001}.

In the reduced-MHD approximation, most of the power is assumed to lie in the $k_\parallel \ll k_\perp$ region. Additionally, the fast waves are ordered out, leaving the only normal modes to be the slow and Alfvén waves. Consequently, the Helmholtz decomposition is not useful in separating out these modes and necessitates decomposing the velocity field along the slow and the Alfvén eigenvectors to analyze their injection rates. Indeed, because both the linear MRI and its linear order parasitic  (Kelvin-Helmholtz) instability are fundamentally incompressible in nature, the turbulence is not expected to produce significant power in fully compressive modes (i.e., polarized along $\hat{k}$), as found in \citet{Bacchini2024}. This, however, is not in contradiction to the result of \citet{Kawazura2022}, i.e., that the slow-to-Alfv\'enic power in the same turbulence is larger than unity.

\bibliographystyle{apj}
\bibliography{main}

\begin{thebibliography}{}
\expandafter\ifx\csname natexlab\endcsname\relax\def\natexlab#1{#1}\fi

\bibitem[{{Akiyama} {et~al.}(2022){Akiyama}, {Alberdi}, {Alef}, {Algaba},
  {Anantua}, {Asada}, {Azulay}, {Bach}, {Baczko}, {Ball}, {Balokovi{\'c}},
  {Barrett}, {Baub{\"o}ck}, {Benson}, {Bintley}, {Blackburn}, {Blundell},
  {Bouman}, {Bower}, {Boyce}, {Bremer}, {Brinkerink}, {Brissenden}, {Britzen},
  {Broderick}, {Broguiere}, {Bronzwaer}, {Bustamante}, {Byun}, {Carlstrom},
  {Ceccobello}, {Chael}, {Chan}, {Chatterjee}, {Chatterjee}, {Chen}, {Chen},
  {Cheng}, {Cho}, {Christian}, {Conroy}, {Conway}, {Cordes}, {Crawford},
  {Crew}, {Cruz-Osorio}, {Cui}, {Davelaar}, {De Laurentis}, {Deane}, {Dempsey},
  {Desvignes}, {Dexter}, {Dhruv}, {Doeleman}, {Dougal}, {Dzib}, {Eatough},
  {Emami}, {Falcke}, {Farah}, {Fish}, {Fomalont}, {Ford}, {Fraga-Encinas},
  {Freeman}, {Friberg}, {Fromm}, {Fuentes}, {Galison}, {Gammie}, {Garc{\'\i}a},
  {Gentaz}, {Georgiev}, {Goddi}, {Gold}, {G{\'o}mez-Ruiz}, {G{\'o}mez}, {Gu},
  {Gurwell}, {Hada}, {Haggard}, {Haworth}, {Hecht}, {Hesper}, {Heumann}, {Ho},
  {Ho}, {Honma}, {Huang}, {Huang}, {Hughes}, {Ikeda}, {Impellizzeri}, {Inoue},
  {Issaoun}, {James}, {Jannuzi}, {Janssen}, {Jeter}, {Jiang},
  {Jim{\'e}nez-Rosales}, {Johnson}, {Jorstad}, {Joshi}, {Jung}, {Karami},
  {Karuppusamy}, {Kawashima}, {Keating}, {Kettenis}, {Kim}, {Kim}, {Kim},
  {Kim}, {Kino}, {Koay}, {Kocherlakota}, {Kofuji}, {Koch}, {Koyama}, {Kramer},
  {Kramer}, {Krichbaum}, {Kuo}, {La Bella}, {Lauer}, {Lee}, {Lee}, {Leung},
  {Levis}, {Li}, {Lico}, {Lindahl}, {Lindqvist}, {Lisakov}, {Liu}, {Liu},
  {Liuzzo}, {Lo}, {Lobanov}, {Loinard}, {Lonsdale}, {Lu}, {Mao}, {Marchili},
  {Markoff}, {Marrone}, {Marscher}, {Mart{\'\i}-Vidal}, {Matsushita},
  {Matthews}, {Medeiros}, {Menten}, {Michalik}, {Mizuno}, {Mizuno}, {Moran},
  {Moriyama}, {Moscibrodzka}, {M{\"u}ller}, {Mus}, {Musoke}, {Myserlis},
  {Nadolski}, {Nagai}, {Nagar}, {Nakamura}, {Narayan}, {Narayanan},
  {Natarajan}, {Nathanail}, {Navarro Fuentes}, {Neilsen}, {Neri}, {Ni},
  {Noutsos}, {Nowak}, {Oh}, {Okino}, {Olivares}, {Ortiz-Le{\'o}n}, {Oyama},
  {{\"O}zel}, {Palumbo}, {Paraschos}, {Park}, {Parsons}, {Patel}, {Pen},
  {Pesce}, {Pi{\'e}tu}, {Plambeck}, {PopStefanija}, {Porth}, {P{\"o}tzl},
  {Prather}, {Preciado-L{\'o}pez}, {Psaltis}, {Pu}, {Ramakrishnan}, {Rao},
  {Rawlings}, {Raymond}, {Rezzolla}, {Ricarte}, {Ripperda}, {Roelofs},
  {Rogers}, {Ros}, {Romero-Ca{\~n}izales}, {Roshanineshat}, {Rottmann}, {Roy},
  {Ruiz}, {Ruszczyk}, {Rygl}, {S{\'a}nchez}, {S{\'a}nchez-Arg{\"u}elles},
  {S{\'a}nchez-Portal}, {Sasada}, {Satapathy}, {Savolainen}, {Schloerb},
  {Schonfeld}, {Schuster}, {Shao}, {Shen}, {Small}, {Sohn}, {SooHoo},
  {Souccar}, {Sun}, {Tazaki}, {Tetarenko}, {Tiede}, {Tilanus}, {Titus},
  {Torne}, {Traianou}, {Trent}, {Trippe}, {Turk}, {van Bemmel}, {van
  Langevelde}, {van Rossum}, {Vos}, {Wagner}, {Ward-Thompson}, {Wardle},
  {Weintroub}, {Wex}, {Wharton}, {Wielgus}, {Wiik}, {Witzel}, {Wondrak},
  {Wong}, {Wu}, {Yamaguchi}, {Yoon}, {Young}, {Young}, {Younsi}, {Yuan},
  {Yuan}, {Zensus}, {Zhang}, {Zhao}, {Zhao}, {Chan}, {Qiu}, {Ressler}, {White},
  \& {EHT Collaboration}}]{EHT2022e}
{Akiyama}, K., {Alberdi}, A., {Alef}, W., {et~al.} 2022, \apjl, 930, L16

\bibitem[{{Bacchini} {et~al.}(2024){Bacchini}, {Zhdankin}, {Gorbunov},
  {Werner}, {Arzamasskiy}, {Begelman}, \& {Uzdensky}}]{Bacchini2024}
{Bacchini}, F., {Zhdankin}, V., {Gorbunov}, E.~A., {et~al.} 2024, arXiv
  e-prints, arXiv:2401.01399

\bibitem[{{Balbus} \& {Hawley}(1991)}]{Balbus1991}
{Balbus}, S.~A., \& {Hawley}, J.~F. 1991, \apj, 376, 214

\bibitem[{{Balbus} \& {Hawley}(1998)}]{Balbus1998}
---. 1998, Reviews of Modern Physics, 70, 1

\bibitem[{{Ball} {et~al.}(2016){Ball}, {{\"O}zel}, {Psaltis}, \&
  {Chan}}]{Ball2016}
{Ball}, D., {{\"O}zel}, F., {Psaltis}, D., \& {Chan}, C.-k. 2016, \apj, 826, 77

\bibitem[{{Ball} {et~al.}(2018){Ball}, {{\"O}zel}, {Psaltis}, {Chan}, \&
  {Sironi}}]{Ball2018}
{Ball}, D., {{\"O}zel}, F., {Psaltis}, D., {Chan}, C.-K., \& {Sironi}, L. 2018,
  \apj, 853, 184

\bibitem[{{Ball} {et~al.}(2019){Ball}, {Sironi}, \& {{\"O}zel}}]{Ball2019}
{Ball}, D., {Sironi}, L., \& {{\"O}zel}, F. 2019, \apj, 884, 57

\bibitem[{{Chael} {et~al.}(2018){Chael}, {Rowan}, {Narayan}, {Johnson}, \&
  {Sironi}}]{Chael2018a}
{Chael}, A., {Rowan}, M., {Narayan}, R., {Johnson}, M., \& {Sironi}, L. 2018,
  \mnras, 478, 5209

\bibitem[{{Cho} \& {Lazarian}(2003)}]{ChoLazarian2003}
{Cho}, J., \& {Lazarian}, A. 2003, \mnras, 345, 325

\bibitem[{{Dexter} {et~al.}(2020){Dexter}, {Jim{\'e}nez-Rosales}, {Ressler},
  {Tchekhovskoy}, {Baub{\"o}ck}, {de Zeeuw}, {Eisenhauer}, {von Fellenberg},
  {Gao}, {Genzel}, {Gillessen}, {Habibi}, {Ott}, {Stadler}, {Straub}, \&
  {Widmann}}]{Dexter2020}
{Dexter}, J., {Jim{\'e}nez-Rosales}, A., {Ressler}, S.~M., {et~al.} 2020,
  \mnras, 494, 4168

\bibitem[{{Event Horizon Telescope Collaboration} {et~al.}(2019){Event Horizon
  Telescope Collaboration}, {Akiyama}, {Alberdi}, {Alef}, {Asada}, {Azulay},
  {Baczko}, {Ball}, {Balokovi{\'c}}, {Barrett}, {Bintley}, {Blackburn},
  {Boland}, {Bouman}, {Bower}, {Bremer}, {Brinkerink}, {Brissenden}, {Britzen},
  {Broderick}, {Broguiere}, {Bronzwaer}, {Byun}, {Carlstrom}, {Chael}, {Chan},
  {Chatterjee}, {Chatterjee}, {Chen}, {Chen}, {Cho}, {Christian}, {Conway},
  {Cordes}, {Crew}, {Cui}, {Davelaar}, {De Laurentis}, {Deane}, {Dempsey},
  {Desvignes}, {Dexter}, {Doeleman}, {Eatough}, {Falcke}, {Fish}, {Fomalont},
  {Fraga-Encinas}, {Friberg}, {Fromm}, {G{\'o}mez}, {Galison}, {Gammie},
  {Garc{\'\i}a}, {Gentaz}, {Georgiev}, {Goddi}, {Gold}, {Gu}, {Gurwell},
  {Hada}, {Hecht}, {Hesper}, {Ho}, {Ho}, {Honma}, {Huang}, {Huang}, {Hughes},
  {Ikeda}, {Inoue}, {Issaoun}, {James}, {Jannuzi}, {Janssen}, {Jeter}, {Jiang},
  {Johnson}, {Jorstad}, {Jung}, {Karami}, {Karuppusamy}, {Kawashima},
  {Keating}, {Kettenis}, {Kim}, {Kim}, {Kim}, {Kino}, {Koay}, {Koch}, {Koyama},
  {Kramer}, {Kramer}, {Krichbaum}, {Kuo}, {Lauer}, {Lee}, {Li}, {Li},
  {Lindqvist}, {Liu}, {Liuzzo}, {Lo}, {Lobanov}, {Loinard}, {Lonsdale}, {Lu},
  {MacDonald}, {Mao}, {Markoff}, {Marrone}, {Marscher}, {Mart{\'\i}-Vidal},
  {Matsushita}, {Matthews}, {Medeiros}, {Menten}, {Mizuno}, {Mizuno}, {Moran},
  {Moriyama}, {Moscibrodzka}, {Mul{\ensuremath{\ddot{}}}ler}, {Nagai}, {Nagar},
  {Nakamura}, {Narayan}, {Narayanan}, {Natarajan}, {Neri}, {Ni}, {Noutsos},
  {Okino}, {Olivares}, {Oyama}, {{\"O}zel}, {Palumbo}, {Patel}, {Pen}, {Pesce},
  {Pi{\'e}tu}, {Plambeck}, {PopStefanija}, {Porth}, {Prather},
  {Preciado-L{\'o}pez}, {Psaltis}, {Pu}, {Ramakrishnan}, {Rao}, {Rawlings},
  {Raymond}, {Rezzolla}, {Ripperda}, {Roelofs}, {Rogers}, {Ros}, {Rose},
  {Roshanineshat}, {Rottmann}, {Roy}, {Ruszczyk}, {Ryan}, {Rygl},
  {S{\'a}nchez}, {S{\'a}nchez-Arguelles}, {Sasada}, {Savolainen}, {Schloerb},
  {Schuster}, {Shao}, {Shen}, {Small}, {Sohn}, {SooHoo}, {Tazaki}, {Tiede},
  {Tilanus}, {Titus}, {Toma}, {Torne}, {Trent}, {Trippe}, {Tsuda}, {van
  Bemmel}, {van Langevelde}, {van Rossum}, {Wagner}, {Wardle}, {Weintroub},
  {Wex}, {Wharton}, {Wielgus}, {Wong}, {Wu}, {Young}, {Young}, {Younsi},
  {Yuan}, {Yuan}, {Zensus}, {Zhao}, {Zhao}, {Zhu}, {Anczarski}, {Baganoff},
  {Eckart}, {Farah}, {Haggard}, {Meyer-Zhao}, {Michalik}, {Nadolski},
  {Neilsen}, {Nishioka}, {Nowak}, {Pradel}, {Primiani}, {Souccar},
  {Vertatschitsch}, {Yamaguchi}, \& {Zhang}}]{EHT2019e}
{Event Horizon Telescope Collaboration}, {Akiyama}, K., {Alberdi}, A., {et~al.}
  2019, \apjl, 875, L5

\bibitem[{{Goldreich} \& {Lynden-Bell}(1965)}]{Goldreich1965}
{Goldreich}, P., \& {Lynden-Bell}, D. 1965, \mnras, 130, 125

\bibitem[{{Goodman} \& {Xu}(1994)}]{GoodmanXu1994}
{Goodman}, J., \& {Xu}, G. 1994, \apj, 432, 213

\bibitem[{{Hawley} {et~al.}(1995){Hawley}, {Gammie}, \& {Balbus}}]{Hawley1995}
{Hawley}, J.~F., {Gammie}, C.~F., \& {Balbus}, S.~A. 1995, \apj, 440, 742

\bibitem[{{Kawazura} {et~al.}(2022){Kawazura}, {Schekochihin}, {Barnes},
  {Dorland}, \& {Balbus}}]{Kawazura2022}
{Kawazura}, Y., {Schekochihin}, A.~A., {Barnes}, M., {Dorland}, W., \&
  {Balbus}, S.~A. 2022, Journal of Plasma Physics, 88, 905880311

\bibitem[{{Kawazura} {et~al.}(2020){Kawazura}, {Schekochihin}, {Barnes},
  {TenBarge}, {Tong}, {Klein}, \& {Dorland}}]{Kawazura2020}
{Kawazura}, Y., {Schekochihin}, A.~A., {Barnes}, M., {et~al.} 2020, Physical
  Review X, 10, 041050

\bibitem[{{Kunz} {et~al.}(2018){Kunz}, {Abel}, {Klein}, \&
  {Schekochihin}}]{Kunz2018}
{Kunz}, M.~W., {Abel}, I.~G., {Klein}, K.~G., \& {Schekochihin}, A.~A. 2018,
  Journal of Plasma Physics, 84, 715840201

\bibitem[{{Lesur} \& {Longaretti}(2011)}]{Lesur2011}
{Lesur}, G., \& {Longaretti}, P.~Y. 2011, \aap, 528, A17

\bibitem[{{Maron} \& {Goldreich}(2001)}]{Maron2001}
{Maron}, J., \& {Goldreich}, P. 2001, \apj, 554, 1175

\bibitem[{{Narayan} {et~al.}(2012){Narayan}, {S{\"A} dowski}, {Penna}, \&
  {Kulkarni}}]{Narayan2012}
{Narayan}, R., {S{\"A} dowski}, A., {Penna}, R.~F., \& {Kulkarni}, A.~K. 2012,
  \mnras, 426, 3241

\bibitem[{{Narayan} \& {Yi}(1995)}]{Narayan1995b}
{Narayan}, R., \& {Yi}, I. 1995, \apj, 452, 710

\bibitem[{{{\"O}zel} {et~al.}(2022){{\"O}zel}, {Psaltis}, \&
  {Younsi}}]{Ozel2022}
{{\"O}zel}, F., {Psaltis}, D., \& {Younsi}, Z. 2022, \apj, 941, 88

\bibitem[{{Pessah}(2010)}]{Pessah2010}
{Pessah}, M.~E. 2010, \apj, 716, 1012

\bibitem[{{Pessah} {et~al.}(2006{\natexlab{a}}){Pessah}, {Chan}, \&
  {Psaltis}}]{Pessah2006}
{Pessah}, M.~E., {Chan}, C.-K., \& {Psaltis}, D. 2006{\natexlab{a}}, \prl, 97,
  221103

\bibitem[{{Pessah} {et~al.}(2006{\natexlab{b}}){Pessah}, {Chan}, \&
  {Psaltis}}]{Pessah2006a}
---. 2006{\natexlab{b}}, \mnras, 372, 183

\bibitem[{{Pessah} \& {Psaltis}(2005)}]{Pessah2005}
{Pessah}, M.~E., \& {Psaltis}, D. 2005, \apj, 628, 879

\bibitem[{{Quataert}(1998)}]{Quataert1998}
{Quataert}, E. 1998, \apj, 500, 978

\bibitem[{{Ressler} {et~al.}(2015){Ressler}, {Tchekhovskoy}, {Quataert},
  {Chandra}, \& {Gammie}}]{Ressler2015}
{Ressler}, S.~M., {Tchekhovskoy}, A., {Quataert}, E., {Chandra}, M., \&
  {Gammie}, C.~F. 2015, \mnras, 454, 1848

\bibitem[{{Rowan} {et~al.}(2019){Rowan}, {Sironi}, \& {Narayan}}]{Rowan2019}
{Rowan}, M.~E., {Sironi}, L., \& {Narayan}, R. 2019, \apj, 873, 2

\bibitem[{{Satapathy} {et~al.}(2023){Satapathy}, {Psaltis}, \&
  {{\"O}zel}}]{Satapathy2023}
{Satapathy}, K., {Psaltis}, D., \& {{\"O}zel}, F. 2023, \apj, 955, 47

\bibitem[{{Schekochihin} {et~al.}(2009){Schekochihin}, {Cowley}, {Dorland},
  {Hammett}, {Howes}, {Quataert}, \& {Tatsuno}}]{Schekochihin2009}
{Schekochihin}, A.~A., {Cowley}, S.~C., {Dorland}, W., {et~al.} 2009, \apjs,
  182, 310

\bibitem[{{Sharma} {et~al.}(2007){Sharma}, {Quataert}, {Hammett}, \&
  {Stone}}]{Sharma2007}
{Sharma}, P., {Quataert}, E., {Hammett}, G.~W., \& {Stone}, J.~M. 2007, \apj,
  667, 714

\bibitem[{{S{\k{a}}dowski} {et~al.}(2017){S{\k{a}}dowski}, {Wielgus},
  {Narayan}, {Abarca}, {McKinney}, \& {Chael}}]{Sadowski2017}
{S{\k{a}}dowski}, A., {Wielgus}, M., {Narayan}, R., {et~al.} 2017, \mnras, 466,
  705

\bibitem[{Stone {et~al.}(2020)Stone, Tomida, White, \& Felker}]{Stone2020}
Stone, J.~M., Tomida, K., White, C.~J., \& Felker, K.~G. 2020, The
  Astrophysical Journal Supplement Series, 249, 4

\bibitem[{{Zhdankin}(2021)}]{Zhdankin2021}
{Zhdankin}, V. 2021, \apj, 922, 172

\end{thebibliography}
\end{document}